\documentclass[twocolumn]{aastex7}
\hypersetup{breaklinks}
\usepackage{comment, todonotes}
\usepackage{savesym, amsmath, amstext}
\usepackage{ulem}

\usepackage[utf8]{inputenc}
\usepackage{footmisc}

\def\asec{\ifmmode^{\prime\prime}\else$^{\prime\prime}$\fi}
\newcommand{\caii}{\ion{Ca}{2}}

\newcommand{\rpirt}{\ensuremath{R'_{\rm IRT}}}

\newcommand{\Msun}{\ensuremath{{M}_{\odot}}}

\newcommand{\prot}{\ensuremath{P_{\mbox{\scriptsize rot}}}}

\newcommand{\ith}{\ensuremath{^{\rm th}}}
\newcommand{\gbr}{\ensuremath{(G_{\rm BP} - G_{\rm RP})}}
\def\amin{\ifmmode^{\prime}\else$^{\prime}$\fi}
\newcommand{\lapprox }{{\lower0.8ex\hbox{$\buildrel <\over\sim$}}}
\newcommand{\gapprox }{{\lower0.8ex\hbox{$\buildrel >\over\sim$}}}

\newcommand{\columbia}{Department of Astronomy, Columbia University, 550 West 120th Street, New York, NY 10027, USA}
\newcommand{\LAB}{Laboratoire d’astrophysique de Bordeaux, Université de Bordeaux, CNRS, B18N, Allée Geoffroy Saint-Hilaire, 33615 Pessac, France}

\bibpunct[; ]{(}{)}{,}{a}{}{;}

\shortauthors{Stafford et al.}
\shorttitle{TESS Rotation Periods in Crowded Star Clusters}

\accepted{January 20, 2026}

\begin{document}

\title{\sc{Measuring Rotation Periods in Crowded Star Clusters with TESS:\\A Proof-of-Concept with NGC~3532 }}

%%%%%%%%%%%%%%%%%%%%%%%%%%%%%%%%%%%%%%%%%%%%%%%%%%%%%%%%%%%%%%%%%%%%%%%%%%%%%%%%%%%%%%%%%%%%%%%%%%%%%%%%%%%%%%%%%%
% Authors

\correspondingauthor{Matthew S.~Stafford}
\email{ms5750@columbia.edu}

\author[0000-0002-6438-8353]{Matthew S. Stafford}
\affiliation{\columbia}
\email{ms5750@columbia.edu}

\author[0000-0002-2792-134X]{Jason Lee Curtis}
\affiliation{\columbia}
\email{jasoncurtis.astro@gmail.com}

\author[0000-0001-7077-3664]{Marcel A.~Ag\"{u}eros}
\affiliation{\columbia},\affiliation{\LAB}
\email{m.agueros@columbia.edu}

%%%%%%%%%%%%%%%%%%%%%%%%%%%%%%%%%%%%%%%%%%%%%%%%%%%%%%%%%%%%%%%%%%%%%%%%%%%%%%%%%%%%%%%%%%%%%%%%%%%%%%%%%%%%%%%%%%
% 250 word limit for AAS Journals, 150 for RNAAS
\begin{abstract}
The Transiting Exoplanet Survey Satellite (TESS) has observed nearly the entire sky, producing full-frame images (FFIs) every 30 min (Cycles 1--2), 10 min (Cycles 3--4), and now 200 s (Cycle 5+), over 27-day sectors. Light curves extracted from FFIs can be used to measure stellar rotation periods (\prot) in nearby open clusters, and are well-suited for studying low-mass stars ($\lapprox$1.2~\Msun) younger than $\lapprox$1~Gyr, whose \prot\ are generally still $\leq$15~days. A challenge to exploiting TESS data fully is its 21$\asec$ pixel size, which can cause strong signals from a source to contaminate the signals of nearby sources in the crowded environments found, e.g., in the more distant and/or richest clusters. We conducted a test with the young ($\approx$350 Myr old), moderately distant (470~pc), and rich open cluster NGC~3532 ($N_\star >$~3000), which has an extensive \prot\ catalog from ground-based photometry, to examine the reliability of \prot\ obtained from TESS data. We recovered 69\% of the literature periods from at least one of the three TESS cycles in which NGC~3532 was observed before any quality analysis. We then used all available TESS data for low-mass members of NGC 3532 and, applying a set of quality cuts that combined information from TESS and from Gaia, measured \prot\ for 885 cluster stars, adding 706 new \prot\ to the existing catalog. We conclude that, when considered with appropriate caution, TESS data for stars in crowded fields can yield reliable \prot\ measurements.
\end{abstract}


%%%%%%%%%%%%%%%%%%%%%%%%%%%%%%%%%%%%%%%%%%%%%%%%%%%%%%%%%%%%%%%%%%%%%%%%%%%%%%%%%%%%%%%%%%%%%%%%%%%%%%%%%%%%%%%%%%
\section{Introduction} \label{sec:second}

\begin{figure*}[!th]
%\centerline{\includegraphics[width=1\textwidth, trim=0.7cm 0.65cm 0.6cm 0.0cm, clip=True]{Crowding_Example-TESS_vs_DSS-EB_Full-2025Oct03.png}}
\centerline{\includegraphics[width=1\textwidth, trim=0.7cm 0.65cm 0.6cm 0.0cm, clip=True]{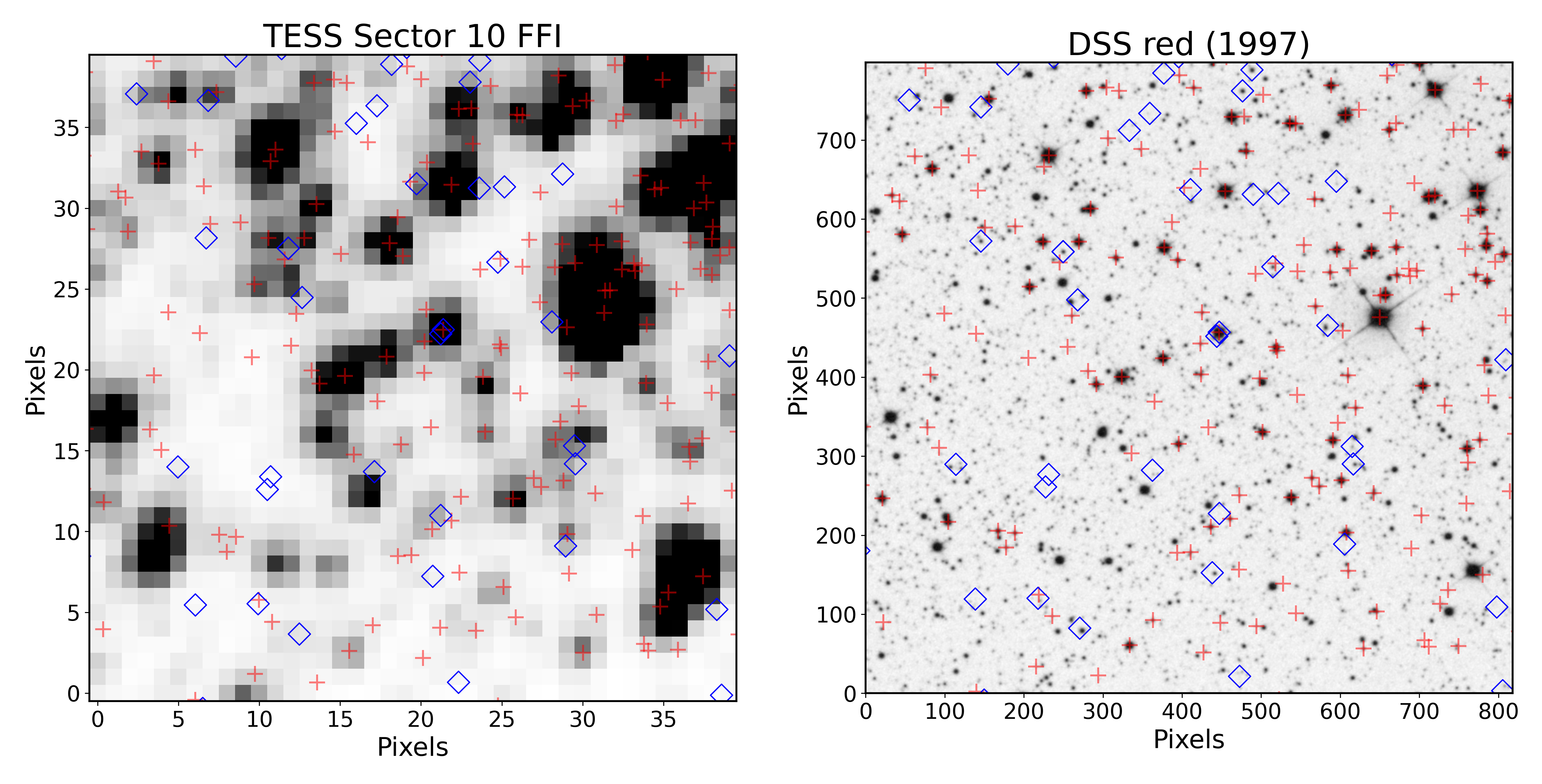}}
\caption{Comparison of a TESS Sector 10 FFI 40$\times$40 pixel cutout (left) and ground-based DSS image (right) for a representative cluster target. The red plus symbols are the 159 \citet{Hunt2023} cluster members in the image, and the blue diamonds are 32 variables included in the Gaia DR3 variability catalog  \citep[][]{GaiaDR3_Variability_method}. Clearly, the TESS pixels are large enough to  include contributions from several sources, including background variables.
\label{fig:DSS}}
\end{figure*}

Unlike its predecessor Kepler \citep[and then K2;][]{borucki2010,howell2014}, NASA's Transiting Exoplanet Survey Satellite \citep[TESS;][]{Ricker2015} is a (nearly) all-sky telescope. TESS data can therefore be used to measure the \prot\ of cool main-sequence stars across much of the sky \citep[e.g.,][]{curtis2019b,anthony2022, newton2022, pass2022, rebull2022, frasca2023, popinchalk2023, petrucci2024, claytor2024, Colman2024}. However, TESS's pixels are large (21$\asec$ vs.~4$\asec$ for Kepler/K2), potentially limiting the usefulness of these data for measuring \prot\ in distant clusters and/or crowded Galactic fields due to photometric contamination from nearby stars (see Figure~\ref{fig:DSS}).

We used the open cluster NGC~3532 (R.A.~= 11$^{\rm h}$06$^{\rm m}$, decl.~= $-$58$^\circ$42$\amin$, J2000) as a test case for determining the robustness of \prot\ measurements made from light curves extracted from TESS FFIs in crowded fields. NGC~3532, observed during TESS Cycles 1, 3, and 5 at the time of this study, is $\approx$350 Myr old, relatively distant (470~pc) and rich, with several thousand members \citep{cummings2018,CG2018, fritz2019_rvs,kounkel2019,Hunt2023}. Furthermore, the cluster is centrally concentrated and located in the Galactic plane.
This makes it an excellent test case for studying the impact on period measurements of other variable cluster members or of the large numbers of variable stars in the Galactic background that may contaminate our TESS light curves. NGC~3532 also has a recent and extensive \prot\ catalog  derived from ground-based observations \citep{fritz2021_periods} with which to compare our TESS-based measurements.

Using TESS full-frame images (FFIs), we constructed light curves for 1358 members of NGC~3532 based on the \citet{Hunt2023} catalog, and generated periodograms to find the most likely TESS-based \prot\ for each observation of these stars. We first tested the usefulness of this automated approach by focusing on the subset of the 1358 members that are in the \citet{fritz2021_periods} catalog, and found that we recovered a \prot\ within 15\% of these authors' value for 182/264 (69\%) stars in at least one TESS cycle. Such a high recovery rate, without any post-processing to verify the quality of the \prot\ measurements, indicated that these TESS data can be used to successfully measure \prot\ in this relatively crowded field.

We therefore undertook a comprehensive analysis of the quality of our measurements to produce an expanded \prot\ catalog for the cluster. By applying various quality constraints to our automatic measurements to remove those we determined to be of low quality, and by examining the impact of e.g., source confusion, we obtained confident \prot\ measurements for 885 members of NGC~3532, 179 of which have a \cite{fritz2021_periods} \prot\ measurement, and 706 of which did not previously have a reported \prot.

Our approach differs from the standard point-spread-function and difference-imaging techniques  used to deblend TESS data and produce light curves for studying, e.g., transiting exoplanets and eclipsing binaries \citep[e.g.,][]{Oelkers2018,CDIPS,Feinstein2019,TGLC,PATHOS}. These typically assume a constant flux for sources neighboring a target, making blending from variable sources a persistent concern.

Our approach also differs from other approaches that address the effects of blending by identifying variable sources that are contaminating the targets' signals. For example, with the package \texttt{TESS\_localize} \citep{Higgins2023}, one can use a frequency-based analysis to localize the source(s) of the variability measured in a given light curve. We focused instead on identifying the likely contaminating sources of the measured variability  based on for example, constructing periodograms across multiple pixels, as was done for Kepler data by \cite{Colman2017}.

We then repeated our comparison with the \prot\ from \citet{fritz2021_periods} catalog, now using final \prot\ from our expanded catalog. The overall agreement between the two sets of measurements is 77\%, and it is as high as 86\% for the \citet{fritz2021_periods} class 1 \prot, which are these authors' most confident measurements. We concluded that our approach, which relies only on data from TESS and from Gaia, can be fruitfully applied to other (relatively) crowded stellar fields, and in particular to clusters that do not have an existing \prot\ catalog. 

This paper is structured as follows: in Section~\ref{sec:cat}, we discuss the \cite{fritz2019_rvs, fritz2021_periods, fritz2021_activity}~studies of NGC~3532, which provided us with the comparison sample of \prot\ with which we tested our TESS-based \prot\ measurements. We also describe the \citet{Hunt2023} NGC~3532 cluster membership catalog, which we used to expand the rotation catalog for the cluster. In Section~\ref{sec:prot}, we discuss our procedure for measuring \prot\ from light curves extracted from TESS FFIs, and describe quality filters applied to these measurements to improve the reliability of the \prot\ samples obtained. In Section~\ref{sec:expand}, we provide the updated \prot\ catalog for NGC~3532 that incorporates the application of our \prot\ measurement technique to the TESS data, and compare our final \prot\ measurements to those of \citet{fritz2021_periods}. We conclude in Section~\ref{sec:con}.

%%%%%%%%%%%%%%%%%%%%%%%%%%%%%%%%%%%%%%%%%%%%%%%%%%%%%%%%%%%%%%%%%%%%%%%%%%%%%%%%%%%%%%%%%%%%%%%%%%%%%%%%%%%%%%%%%%

\section{The Existing Rotation and Membership Catalogs for NGC~3532}\label{sec:cat}

\subsection{The D.~J.~Fritzewski et al.~Studies of the Cluster}\label{sec:past}

In the first in a series of three papers, \cite{fritz2019_rvs} produced a membership catalog for NGC~3532, supplementing Gaia Data Release 2 \citep[DR2;][]{gaia2018} proper motions %($\mu$) 
with radial velocity measurements from the literature and from their own observations, and identifying 660 high-confidence members. 

\cite{fritz2021_periods} then imaged NGC~3532 in the $V$ and $I_c$ bands over the course of $\approx$40 consecutive nights with the Yale 1-m telescope at the Cerro Tololo Inter-American Observatory, Chile. \cite{fritz2021_periods} used these data to identify 165 rotators among the radial velocity cluster members, and an additional 11 rotators among Gaia DR2 proper motion members for which \cite{fritz2019_rvs} did not obtain radial velocities. 

\begin{figure}[!t]
%\centerline{\includegraphics[width=\columnwidth, trim=0.65cm 0.6cm 0.6cm 0.cm, clip=True]{Hunt_3532_CMD.png}}
\centerline{\includegraphics[width=\columnwidth, trim=0.65cm 0.6cm 0.6cm 0.cm, clip=True]{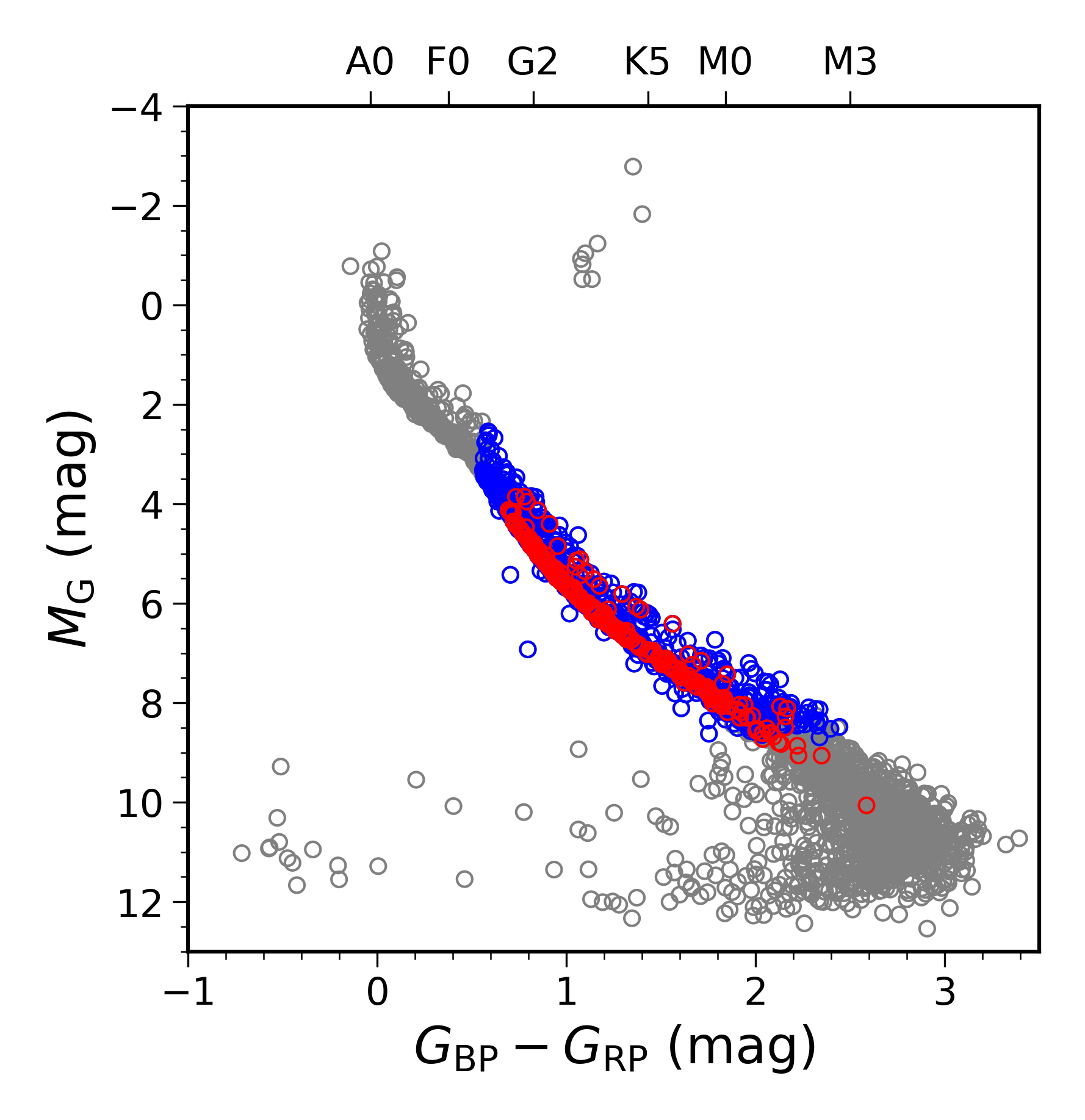}}
\caption{Gaia CMD for NGC~3532 based on the \citet{Hunt2023} catalog of 3414 members. The absolute magnitudes were calculated using inverted Gaia~DR3 parallaxes; we applied a reddening $A_V$ = 0.093~mag \citep{cummings2018}. The red circles are the 276 stars in the \citet{Hunt2023} catalog with \cite{fritz2021_periods} \prot\ measurements.  The blue circles are the 1358 low-mass, main-sequence cluster members for which we have TESS light curves and that met our selection criteria for making new \prot\ measurements.
\label{fig:cmd}}
\end{figure} 

Finally, \cite{fritz2021_activity} used spectra of the \caii\ infrared triplet (IRT) obtained with the AAOmega spectrograph on the 3.9-m Anglo-Australian Telescope, Australia, to measure chromospheric emission ratios \rpirt\ for 454 cluster members. These authors derived an empirical rotation--activity relationship with these measurements and \prot\ from \citet{fritz2021_periods}, and then predicted \prot\ based on \rpirt\ measurements for stars for which \citet{fritz2021_periods} had not measured photometric \prot.  \cite{fritz2021_activity} used the predicted periods as priors when searching those stars' light curves for \prot, and in this manner reported \prot\ for 103 additional NGC~3532 members.\footnote{While the \citet{fritz2021_periods,fritz2021_activity} papers stated that \prot\ for 113 additional stars were obtained using activity measurements in this manner, the published catalog only includes 103 such stars.}

These \citet{fritz2021_activity} \prot\ were included in the \citet{fritz2021_periods} sample. This sample therefore includes 279 \prot\ for F, G, K, and M members of NGC~3532, divided into four classes that reflect their quality and how they were derived. Class 1 \prot\ are measured from light curves with signals clearly evident with visual inspection (93 \prot), class 2 are photometrically derived from light curves with less clear signals (66 \prot), and class 3 are activity-informed (103 \prot); in addition, periods that are possible aliases were labeled class 0 (17 \prot).

The majority of these rotators fall on the sequence of (relatively) slowly rotating stars in the color--period diagram (CPD) for the cluster, extending from $\approx$4~days for the late F stars to $\approx$15~days for the late K/early M stars \citep[see figure 6 in][]{fritz2021_periods}. This is consistent with expectations from younger and older clusters that show these slow-rotating sequences across this same mass range \citep[e.g., at $\approx$120~Myr in the Pleiades, and at $\approx$700 Myr in the Hyades, Praesepe, and Coma Berenices;][]{Barnes2003, Barnes2007, covey2016,rebull2016,douglas2019,Rampalli2021,agueros2025}.

\subsection{The \citet{Hunt2023} \\ Membership Catalog}\label{sec:Hunt_cat}

Even in the case of historical clusters such as NGC~3532, known by Western astronomers since the 18$^{\rm th}$ century and greatly appreciated by no less a figure than John Herschel,\footnote{Herschel described NGC~3532 as ``a glorious cluster of immense magnitude'' in his observing logs from his time in South Africa (1834--1838). The only other objects for which he used the adjective glorious are $\omega$ Cen and 47 Tuc \citep{herschel1847}.} Gaia data have proved extremely useful in improving studies of membership. By applying the \texttt{HDBSCAN} algorithm \citep{Campello2013,hdbscan2017} to the Gaia Data Release 3 \citep[DR3;][]{gaiadr3}, \citet{Hunt2023} found that NGC~3532 has 3414 members.\footnote{\texttt{HDBSCAN}, which is based on \texttt{DBSCAN}, finds clusters based on the density of data points.
\citet{Hunt2021} tested various clustering algorithms and argued that the most effective method for identifying cluster members in Gaia data is to use \texttt{HDBSCAN} with some post-processing to remove contaminants.}$^,$\footnote{Gaia DR3 improved on DR2 parallax precisions by 30\% and proper motion precisions by $\times$2 \citep{gaiadr3}.} The \citet{Hunt2023} catalog is shown in the color--magnitude diagram (CMD) in Figure~\ref{fig:cmd}.

The \cite{Hunt2023} membership catalog for NGC~3532 includes 627 of the 660 high-confidence members from \cite{fritz2019_rvs}. The overlap between catalogs is even greater for the rotators identified by Fritzewski et al. Of the 279 stars in the \citet{fritz2021_periods} \prot\ catalog, 276 are included in the \citet{Hunt2023} membership catalog; we highlight these stars in Figure~\ref{fig:cmd}, and adopt the \citet{Hunt2023} catalog for this study.

\section{Measuring \prot\ for NGC~3532 members}\label{sec:prot}

\subsection{Assembling and Analyzing TESS Light Curves for 1358 Cluster Stars}

We restricted our search for TESS data to the \citet{Hunt2023} stars that met the following criteria: $\gbr > 0.6$~mag, to remove stars lacking a convective outer envelope; $G > 9$~mag, to remove giants; and $G < 17$, to limit our analysis to stars bright enough that their TESS light curves are usable. The overlap between the \citet{Hunt2023} and \citet{fritz2021_periods} catalog described in Section~\ref{sec:Hunt_cat} (276 stars) includes 12 stars with $G>17$~mag that did not meet our criteria for \prot\ measurement. This left us with 264 stars for which we could compare our \prot\ measurements to that from \citet{fritz2021_periods}.

Of the 1358 stars in the complete \citet{Hunt2023} catalog that met these criteria, 1036 were observed in Cycles 1, 3, and 5, while the remaining 322 were observed in two of the three cycles (see Table~\ref{table:TESS}; these 1358 stars are highlighted in Figure~\ref{fig:cmd}). Of these stars, 391 have a $G<17$ mag cluster member within 1\amin, and 789 have one within 2\amin, underscoring the potential for source confusion in this sample.

\begin{deluxetable}{lcccc}
\centering 
\tabletypesize{\footnotesize}

\tablecaption{TESS observations of NGC~3532
 \label{table:TESS}}

\tablehead{
\colhead{} &
\colhead{Cadence} &
\colhead{Sectors} &
\colhead{Date} &
\colhead{\# of}\\[-0.1in]
\colhead{} &
\colhead{} &
\colhead{} &
\colhead{} &
\colhead{Stars\tablenotemark{a}}
}

\startdata
Cycle 1 & 30 min &  10, 11 & 2019 March 26$-$May 20 &  1357 \\
Cycle 3 & 10 min & 36, 37 & 2021 March 7$-$April 28 &  1054 \\
Cycle 5 & 200 s & 63, 64 & 2023 March 10$-$May 4 &  1341 \\
\enddata

\tablenotetext{a}{The numbers quoted here are for the low-mass stars that met our selection criteria. In total, 1036 such stars were observed in all three cycles, while 322 were observed in two of the three cycles.}

\vspace{-.2cm}
\end{deluxetable}

We first downloaded 40$\times$40 pixel FFI cutouts \citep{TESSdata} with \texttt{TESScut} \citep{brasseur2019} for our 1358 targets. We then used Causal Pixel Modeling \citep[CPM;][]{wang2016}, as implemented in the Python package \texttt{unpopular} \citep{Hattori2022}, to produce light curves. \texttt{unpopular} models and subtracts time-varying systematics based on the behavior of pixels outside an exclusion region centered on the target (the central 9$\times$9 pixels in our case). 

The increasing observing cadence between Cycles 1, 3, and 5 raises the photometric noise, which can suppress the periodogram power for otherwise equivalent signals. For this reason and to limit data volume, we binned all the light curves down to 30 min. 

When stars were observed in consecutive sectors (common in Cycles 1 and 5), we joined the light curves, thereby extending the temporal coverage from 27 to 54~days, and treated them as a single light curve. While these longer-baseline light curves open the possibility of measuring longer \prot, we did not expect a significant number of NGC~3532 stars to have \prot~$>$~15~days because of the cluster's relatively young age. Of the 279 stars with a measured \prot\ in the \cite{fritz2021_periods} catalog, only 11 (4\%) have $\prot > 15$~days. Moreover, none of these were categorized as class 1 periods by \cite{fritz2021_periods}: one is classified as class 2, and the remaining 10 are class 3 (i.e., activity-informed), indicating that these are not among the highest-confidence \prot\ measured by these authors.\footnote{Although we did not expect to recover accurate \prot\ for these 11 stars, we included them in our later comparison of our results and those of \cite{fritz2021_periods} to assess the effectiveness of our method in removing inaccurate or unreliable \prot\ for slow rotators.} 

Furthermore, using light curves produced by stitching together data from multiple TESS sectors to measure longer \prot\ for a star is challenging \citep[e.g.,][]{anthony2022}. New approaches may eventually allow us to extend our investigation to longer \prot\ \citep[e.g.,][]{claytor2024, hattori2025}, but for this study we discounted any $>$15~days \prot\ we measured.

\begin{figure*}[!th]
%\centerline{\includegraphics[width=1\textwidth, trim=0.65cm 0.6cm 0.6cm 0.cm, clip=True]{CPD_Before_After.png}}
\centerline{\includegraphics[width=1\textwidth, trim=0.65cm 0.6cm 0.6cm 0.cm, clip=True]{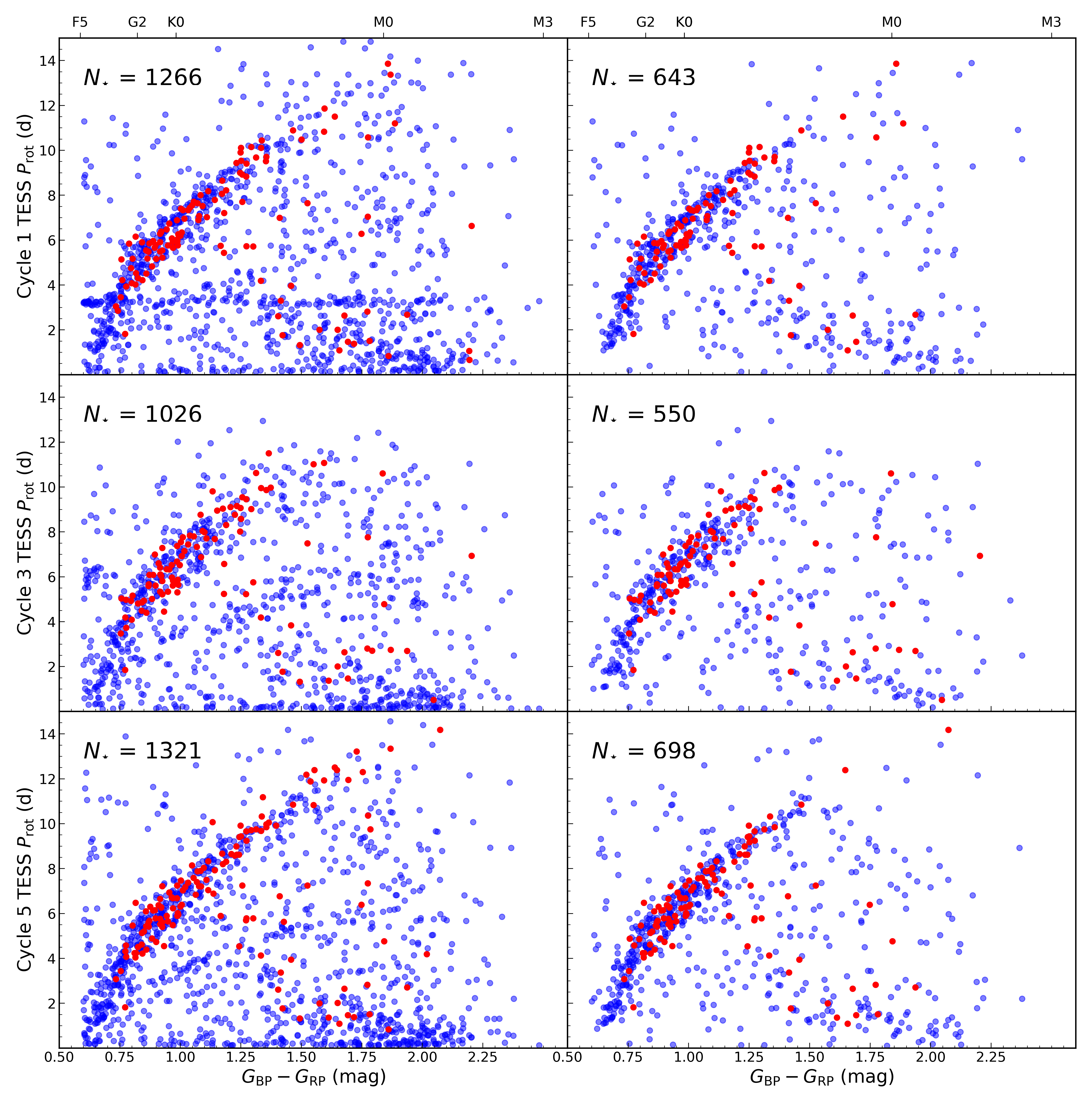}}
\caption{CPDs for NGC~3532 low-mass members with TESS light curves for each TESS cycle in which they were observed before (left) and after (right) our analysis removing erroneous measurements. As in Figure~\ref{fig:cmd}, blue circles represent our targets for \prot\ measurement with TESS; red circles represent stars with a \prot\ measurement that agrees with that from \cite{fritz2021_periods} within 15\%. In each CPD, we omit stars for which we measured $\prot > 15$~days.} 
\label{fig:before_after_CPD}
\end{figure*}

We used \texttt{astropy.timeseries} to calculate Lomb–Scargle (LS) periodograms \citep{Lomb1976, Scargle1982, press1989} with test periods spaced logarithmically between 0.1 and 30~days. As discussed in Section~\ref{sec:har}, we produced both one- and two-term LS periodograms, with the latter used to search for and correct harmonic \prot\ measurements 
\citep[as done by][]{Colman2024}. We recorded the period and power for the top two peaks in our periodograms for each available cycle for the 1358 NGC~3532 stars we considered. For 182/264 (69\%) stars in the \citet{fritz2021_periods} catalog for which we recorded periods at this step, the period associated with the primary periodogram peak for at least one TESS cycle agreed with the \prot\ from \citet{fritz2021_periods} within 15\%. Following this initial comparison, we analyzed our measured \prot\ to develop a method to identify which of our measurements were robust and to remove those that were spurious.

In the left column of Figure~\ref{fig:before_after_CPD}, we show CPDs for the three TESS cycles using the automatically measured TESS \prot\ and Gaia \gbr\ colors. In each CPD, there is, as expected, a densely populated sequence of slowly rotating stars, the position of which is related to the cluster's age. Also visible are clearly erroneous measurements due to e.g., noise, half-period harmonics, and systematic errors, such as the population of 3~day rotators in the top left panel, for Cycle 1. Still, the presence of a well-defined period sequence in the CPDs for all three cycles before \emph{any} assessment of measurement quality demonstrates the potential to retrieve accurate \prot\ measurements.

To identify cases where the TESS-derived \prot\ are inaccurate, we examined the FFIs, light curves extracted from the pixel hosting a given target and from pixels in surrounding annuli, the associated periodograms, and characteristics of neighboring targets (e.g., $G$ magnitude, separation on the sky, \prot). In the remainder of this section, we will describe the several ways we removed spurious measurements from our sample.

\subsection{Removing \prot\ Measurements That Result From Systematics or That Are Marginal Detections}

\begin{figure*}[!th]
%\centerline{\includegraphics[width=1\textwidth, trim=0.6cm 0.65cm 0.6cm 0.0cm, clip=True]{Power_Threshold.png}}
\centerline{\includegraphics[width=1\textwidth, trim=0.6cm 0.65cm 0.6cm 0.0cm, clip=True]{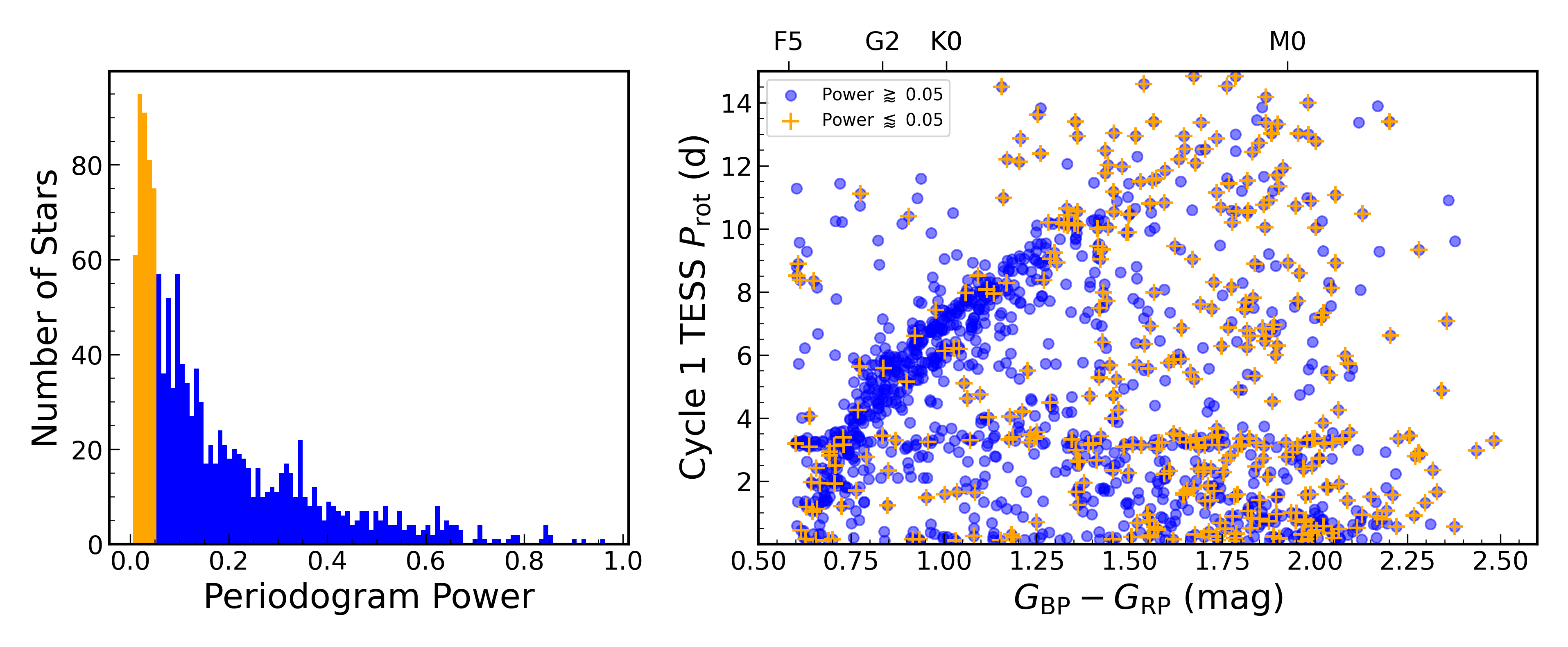}}
\caption{The impact on the Cycle 1 \prot\ distribution of requiring a minimum power when validating \prot\ measurements. The left panel is a histogram of the periodogram powers for the Cycle 1 light curves; the orange bins correspond to powers below the 30\ith\ percentile for this dataset (0.05 in this case). The right panel is the full distribution of Cycle 1 \prot, with the values drawn from measurements with a power below the 30\ith\ percentile indicated with orange crosses. Those low-power measurements include many outliers, such as the K stars with \prot\ $>$ 10 days or $<$1 day, and other problematic measurements, such as the strip of $\approx$3 day \prot\ across the observed color range. 
\label{fig:power_threshold}}
\end{figure*}

Our first step was to set a periodogram power threshold for the \prot\ measurements from each TESS cycle, thereby removing weak signals or otherwise low-quality \prot\ measurements from our sample. A 30\ith\ percentile threshold was applied to all measurements for a given cycle, corresponding to powers of $\approx$0.05, $\approx$0.04, and $\approx$0.03, for Cycles 1, 3, and 5, respectively. This threshold was chosen to identify and remove the peak of low power measurements in the periodogram power histogram for each cycle (see the left panel of Figure~\ref{fig:power_threshold}, for Cycle 1 measurements). The right panel of Figure~\ref{fig:power_threshold} shows the CPD with the low-power \prot\ highlighted. This power threshold identified stars across the full range of \gbr\ and \prot, but especially later type stars that tend to be fainter and include some of the slowest rotators in our sample. % in coeval populations.
Applying this minimum power requirement, we identified 377, 308, and 395 \prot\ measurements for removal in the three cycles.

\begin{figure}
\begin{center}
\includegraphics[width=\columnwidth, trim=0.5cm 0.6cm 0.3cm 0.0cm, clip=True]{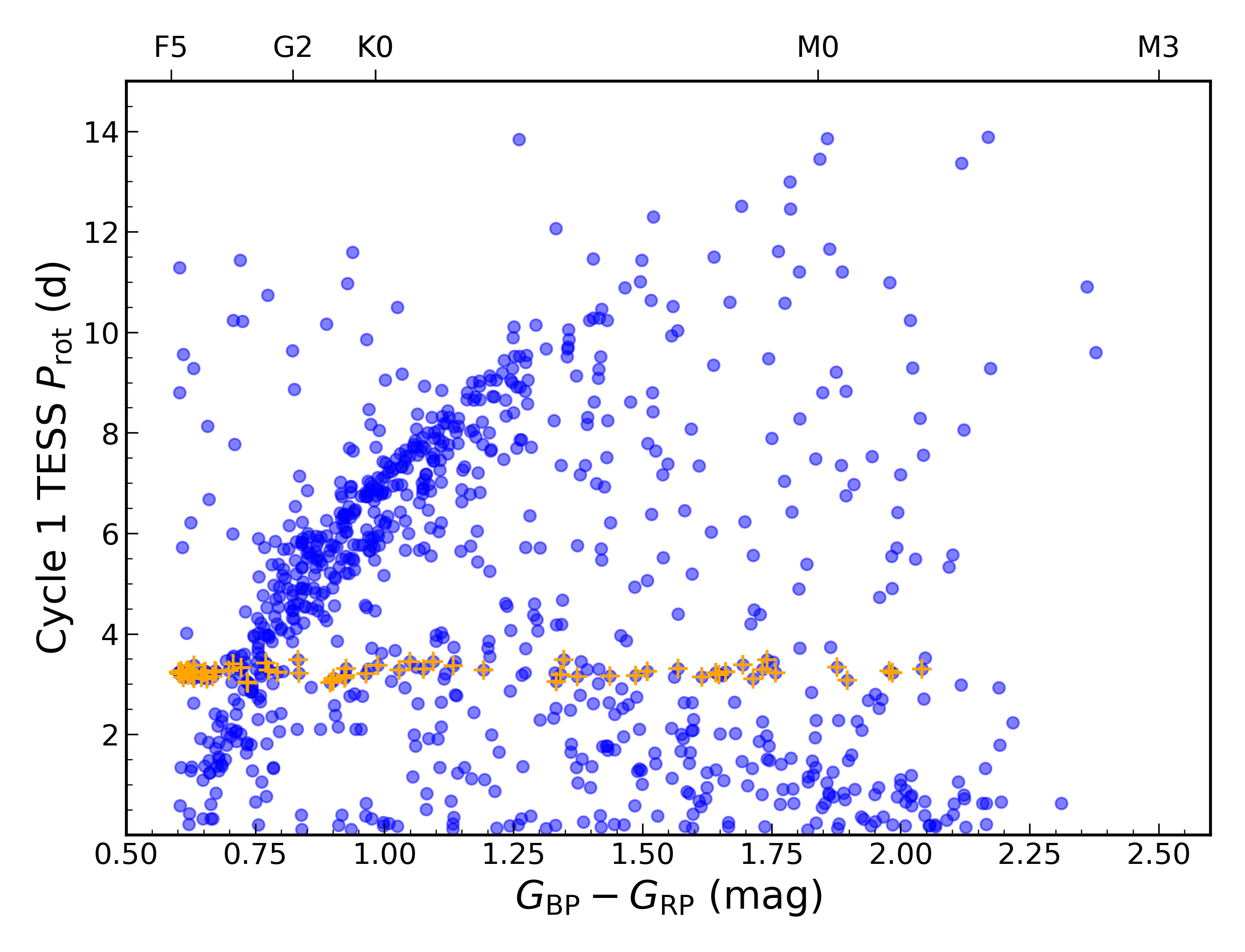}
\caption{TESS Cycle 1 CPD depicting our identification of \prot\ 
measurements that are the result of systematic errors. Measurements of \prot\ $<$~15~days with periodograms powers greater than the 30$^{\rm th}$ percentile power for this cycle are plotted as blue circles. \prot\ measurements ranging from 3.0 to 3.5 days with periodogram powers $<$0.1 are marked with orange crosses and are likely the result of an (unspecified) systematic error in the Cycle 1 data. Setting a \prot\ range and raising the minimum power required for consideration allows us to remove a number of unreliable measurements missed with the cut illustrated in Figure~\ref{fig:power_threshold}.
\label{fig:systematic}}
\end{center}
\end{figure}

The CPDs in the left column of Figure~\ref{fig:before_after_CPD} suggested that there are \prot\ measurements that are the result of artificial signals in the TESS data. This is most evident in the top left panel, the CPD for Cycle 1, where there are a large number of \prot\ between 3.0 and 3.5~days across the full range of color in our sample. Only some of these \prot\ measurements are removed when applying the power requirement described above. The cause of this systematic error is not clear and affects stars distributed across the entire cluster. The periodograms produced from the affected light curves, however, all share a narrow primary peak at similar frequencies. We were able to identify the affected stars by selecting those \prot\ measurements from Cycle 1 between 3.0 and 3.5~days with periodogram powers of $<$0.1. In this manner, we removed 68 \prot\ that were not identified by the previously applied minimum power cut (see Figure~\ref{fig:systematic}).

There is another example of this kind of contamination in the Cycle 3 CPD. In the middle-left panel of Figure~\ref{fig:before_after_CPD}, there is an unexpected clump of \prot\ measurements from 5 to 7~days for stars with $\gbr <0.75$ mag. We removed these erroneous \prot\ measurements in the same fashion, by selecting \prot\ measurements from Cycle 3 between 5.0 and 7.0~days with periodogram powers of $<$0.1, this time also constraining by \gbr. With these criteria, we identified an additional 10 Cycle 3 \prot\ measurements for removal from our sample.

Having access to multiple cycles of TESS data for the same stars also allowed us to identify likely unreliable short \prot\ measurements. We identified the stars for which we measured a $\prot<1$ day that had a \prot\ from another cycle that was slower by at least 20\%. The relatively easy detectability of rapid periodic signals makes it unlikely that a reliable $<$1 day \prot\ measurement would disappear between TESS cycles. We identified 15, 15, and 31 of these rapid \prot\ for removal in our samples for Cycles 1, 3, and 5, respectively.

%%%%%%%%%%%%%%%%%%%%%%%%%%%%%%%%%%%%%%%%%%%%%%%%%%%%%%%%%%%%%%%%%%%%%%%%%%%%%%
\subsection{Identifying Cases of Source Confusion}

The potential blending of signals from neighboring sources can be a source of inaccurate \prot\ measurements when using TESS data for crowded fields. Since NGC~3532 is a young and rich cluster, we expected to detect many variable members, some of which may be near enough to other members to contaminate their light curves. The cluster's location also raises the possibility that the Galactic background may provide many variable stars that can contaminate the TESS light curves.

In cases of source confusion, improperly assigned period measurements will often have high periodogram powers and appear to be robust. These measurements can therefore only be identified as erroneous using an analysis of nearby sources. We performed this analysis using two catalogs: our period measurements for NGC~3532, and the Gaia DR3 variability catalog  \citep[GVC;][]{GaiaDR3_Variability_method}, which includes classifications based on 34 months of Gaia data for nearly 10$^7$ stars.

To illustrate the magnitude of the problem caused by variables in the Galactic background, we searched the GVC for stars within 90\amin\ of the center of NGC~3532, an area that contains 95\% of the cluster's membership. We found 4697 entries: 12 Cepheids \citep{GaiaDR3_Variability_cepheids},
195 RR Lyrae \citep{GaiaDR3_Variability_rrlyrae},
3965 eclipsing binaries \citep[EBs;][]{GaiaDR3_Variability_ebs}, and 525 sources showing short-term variability on timescales $\lesssim$1 day \citep{GaiaDR2_Variability_short}.

\newpage
\subsubsection{An Example of Contamination by \\ a Background Variable}

\begin{figure}
\begin{center}
\includegraphics[width=\columnwidth, trim=0.5cm 0.6cm 0.3cm 0.0cm, clip=True]{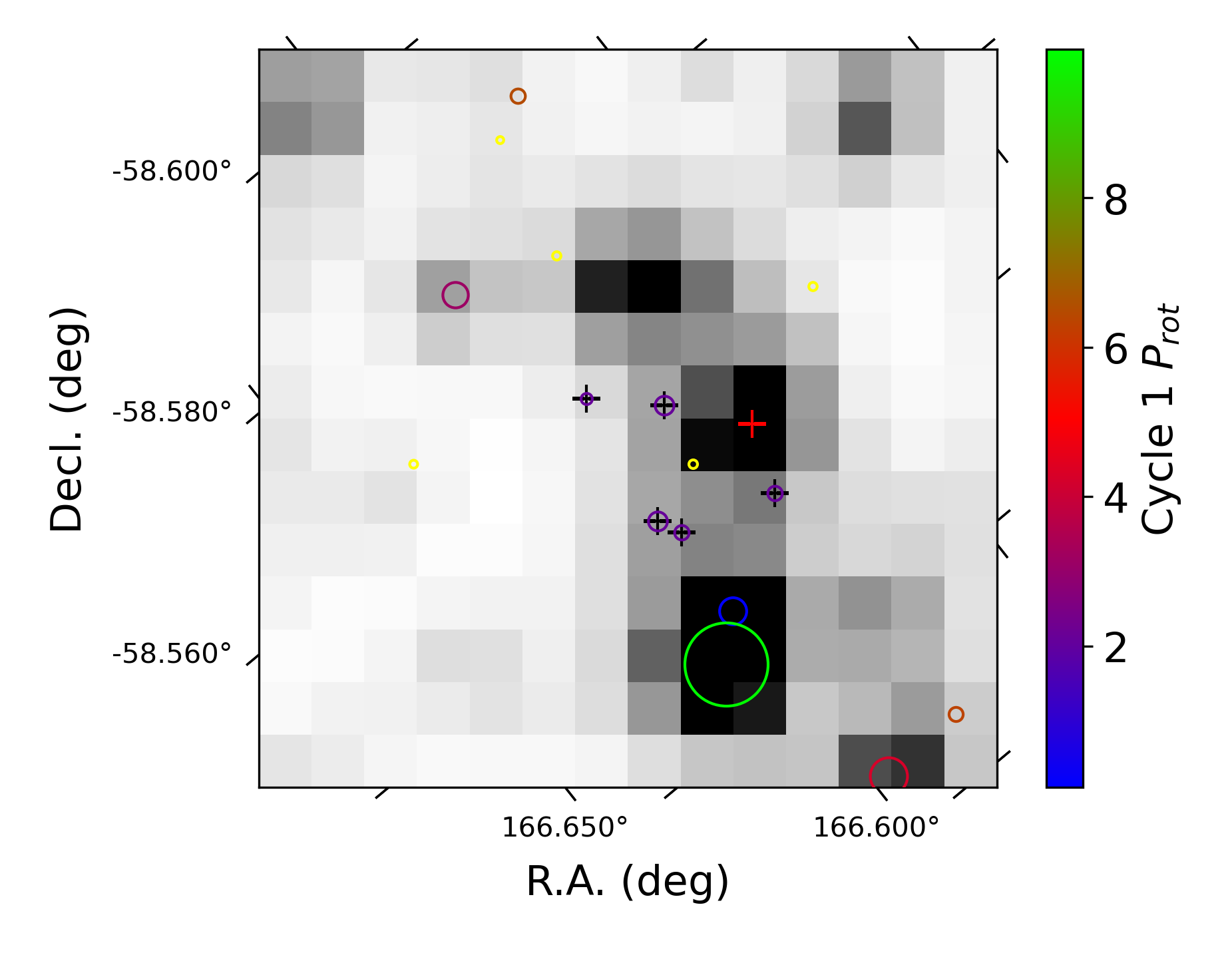}
\caption{14$\times$14 pixel FFI cutout showing the Cycle 1 field containing the Cepheid \object{Gaia DR3 5340164256296564736}, indicated by the red cross. The circles are NGC~3532 members; the size of the circle is related to the star's $G$ magnitude, with larger circles being brighter. The colors of the circles are based on the \prot\ we measured: yellow circles are cluster members for which we do not measure any \prot\ because they are too faint, while the black crosses are members for which we measured a 2.1~day \prot\ that is inaccurate and caused by contamination from the Cepheid. This example demonstrates how bright and rapidly varying sources can prevent accurate \prot\ measurement for nearby targets when using TESS data. 
\label{f:Blending_FFI}}
\end{center}
\end{figure}

A striking example of the impact of contamination by a background GVC source is the set of $\approx$2.1 day periods we measured in Cycle 1 for five cluster members ranging from $G = 12.4 - 14.4$~mag and within about 1\amin\ of each other on the sky (see Figure~\ref{f:Blending_FFI} for an FFI cutout of the field). Four of these five stars have \prot\ measured by \citet{fritz2021_periods}, and these literature measurements are 2--4 times the length of the period we measured, which further raised our suspicions. 

\begin{figure*}[!]
%\centerline{\includegraphics[width=1\textwidth,trim=0.6cm 0.6cm 0.6cm 0.cm, clip=True]{LC_Grid.png}}
\centerline{\includegraphics[width=1\textwidth,trim=0.6cm 0.6cm 0.6cm 0.cm, clip=True]{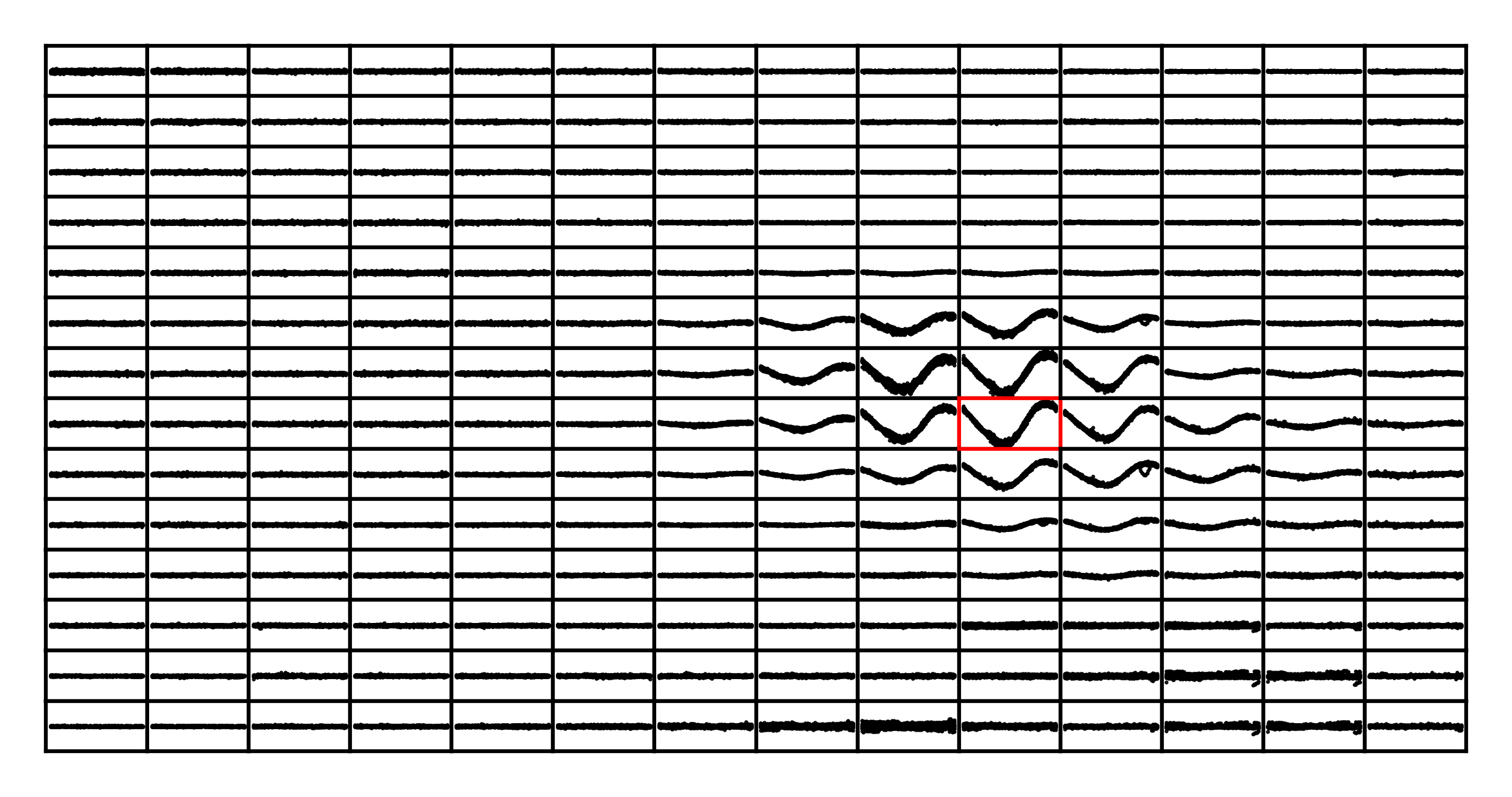}}
\caption{Light curves produced for each pixel in Figure~\ref{f:Blending_FFI} phase-folded on a period of 2.10 days, the period of the classical Cepheid \object{Gaia DR3 5340164256296564736}. The light curve for the pixel hosting the Cepheid is highlighted with a red frame. The 2.10 day signal is dominant in the light curve extracted from that pixel, as expected, but also in those extracted from many surrounding pixels. The y-axes of these plots share a flux scale, demonstrating the relative strength of the 2.10 day signal.}
\label{fig:lc_grid}
\vspace{-0.5cm}
\end{figure*}

\begin{figure*}[!]
%\centerline{\includegraphics[width=1\textwidth, trim=0.6cm 0.6cm 0.6cm 0.cm, clip=True]{PG_Grid.png}}
\centerline{\includegraphics[width=1\textwidth, trim=0.6cm 0.6cm 0.6cm 0.cm, clip=True]{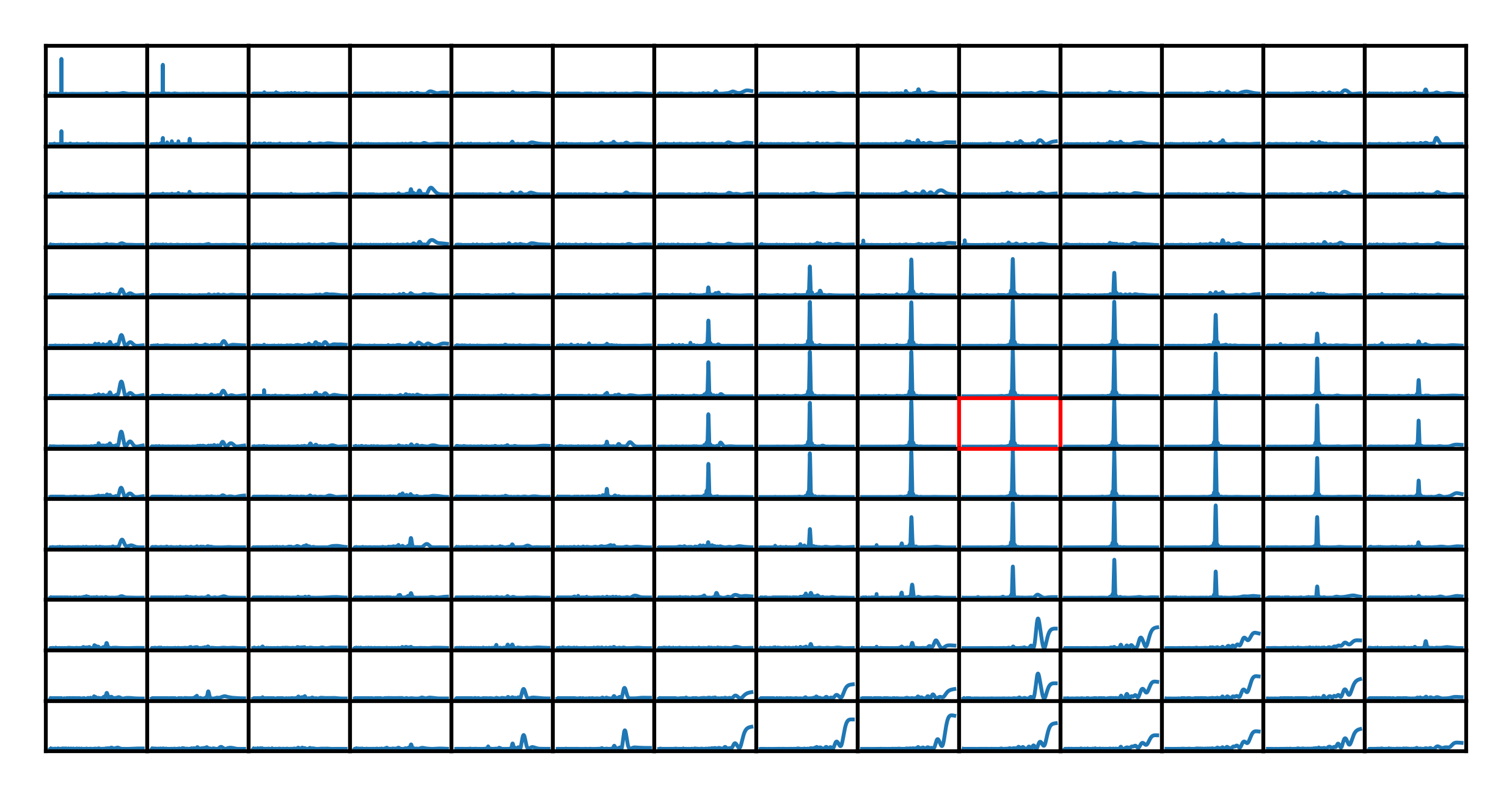}}
\caption{Periodograms produced for the light curves shown in Figure~\ref{fig:lc_grid}. X-axes are logarithmically spaced periods and y-axes are periodogram powers. The periodogram for the pixel hosting \object{Gaia DR3 5340164256296564736} is highlighted with a red frame. The impact of this bright Cepheid on measurements from neighboring pixels is very evident, with its 2.10 day signal being the strongest signal measured in periodograms corresponding to pixels as far removed as $\approx$1$\amin$.}
\label{fig:pg_grid}
\end{figure*}

For each of the pixels shown in Figure~\ref{f:Blending_FFI}, we produced a phase-folded light curve (see Figure~\ref{fig:lc_grid}) and a periodogram (see Figure~\ref{fig:pg_grid}). Examining the FFI and the corresponding light curves and periodograms, it became apparent that the likely source of the periodic signal measured for these cluster members is \object{Gaia DR3 5340164256296564736}, which is significantly brighter than our targets, with $G \approx 10.1$ mag. The GVC lists this star as a classical Cepheid, with a period of 2.1047~days and a $G$ amplitude of 0.13 mag; the periodic signal we measure is caused by pulsations.\footnote{Although older studies included this star, designated as ``NGC 3532 147'' in SIMBAD, as a member of NGC~3532 \citep{Koelbloed1959, Fernandez1980}, \object{Gaia DR3 5340164256296564736} has been omitted from the recent Gaia-based catalogs \citep{kounkel2020, Hunt2023} as it is at a distance of 2.2~kpc.} Due to the star's relative brightness and the rapid periodicity of its variability, its period is recoverable in light curves extracted from pixels several pixels removed from the star's position, corresponding to a separation on the sky $>$1\amin.

This example illustrates how bright and rapidly variable sources can impact \prot\ measurements for nearby stars.

\subsubsection{Mitigating Source Confusion}

For each TESS observation of our 1358 cluster targets, we checked the GVC for sources within 2\amin\ with periods (or harmonics/subharmonics) within 10\% of the \prot\ we measured. This resulted in the removal of 94, 85, and 115 \prot\ that were not removed in previous steps from our samples for Cycles 1, 3, and 5, respectively. Unsurprisingly, given their relative frequency in the GVC, nearly all of the measurements we removed had periods matching that of a nearby EB.

We performed a similar search for possible contamination due to other NGC~3532 members. We first expanded the sample of members we considered as potential contaminants by constructing light curves for 389 stars in the \cite{Hunt2023} membership catalog that did not meet our initial criteria for \prot\ measurement, but whose TESS data are usable (i.e., stars with $\gbr<0.6$~mag or $G<9$), and measured their variability.\footnote{We use ``variability" here to recognize that any periodicity we measure in these light curves is unlikely to be due to rotation.} We then had a sample of 1747 stars to use for this search. 

For each TESS observation of our 1358 targets, we then checked whether any of the other 1746 members were within 2\amin\ on the sky and had a variability measurement within 1\% of our target's \prot. We removed any such \prot\ from our sample if the corresponding periodogram power was less than or equal to the power of the measurement associated with the potential contaminating signal. 

In addition, for stars with \prot\ measured in multiple cycles, we checked the agreement between the \prot\ measurements across the cycles. If in a given cycle, the \prot\ was within 5\% of one flagged as due to contamination in a different cycle, we removed that \prot\ from our sample as well. This resulted in the removal of 69, 58, and 82 \prot\ that were not removed by previous steps from our samples for Cycles 1, 3, and 5, respectively.

%%%%%%%%%%%%%%%%%%%%%%%%%%%%%%%%%%%%%%%%%%%%%%%%%%%%%%%%%%%%%%%%%%%%%%%%%%%%%%%%%%%%%%%%%%%%%%%%%%%%%%%%%

\clearpage
%%%%%%%%%%%%%%%%%%%%%%%%%%%%%%%%%%%%%%%%%%%%%%%%%%%%%%%%%%%%%%%%%%%%%%%%%%%%%%%%%%%%%%%%%%%%%%%%%%%%%%%%%

\begin{figure}
\begin{center}
\includegraphics[width=\columnwidth, trim=0.5cm 0.6cm 0.3cm 0.0cm, clip=True]{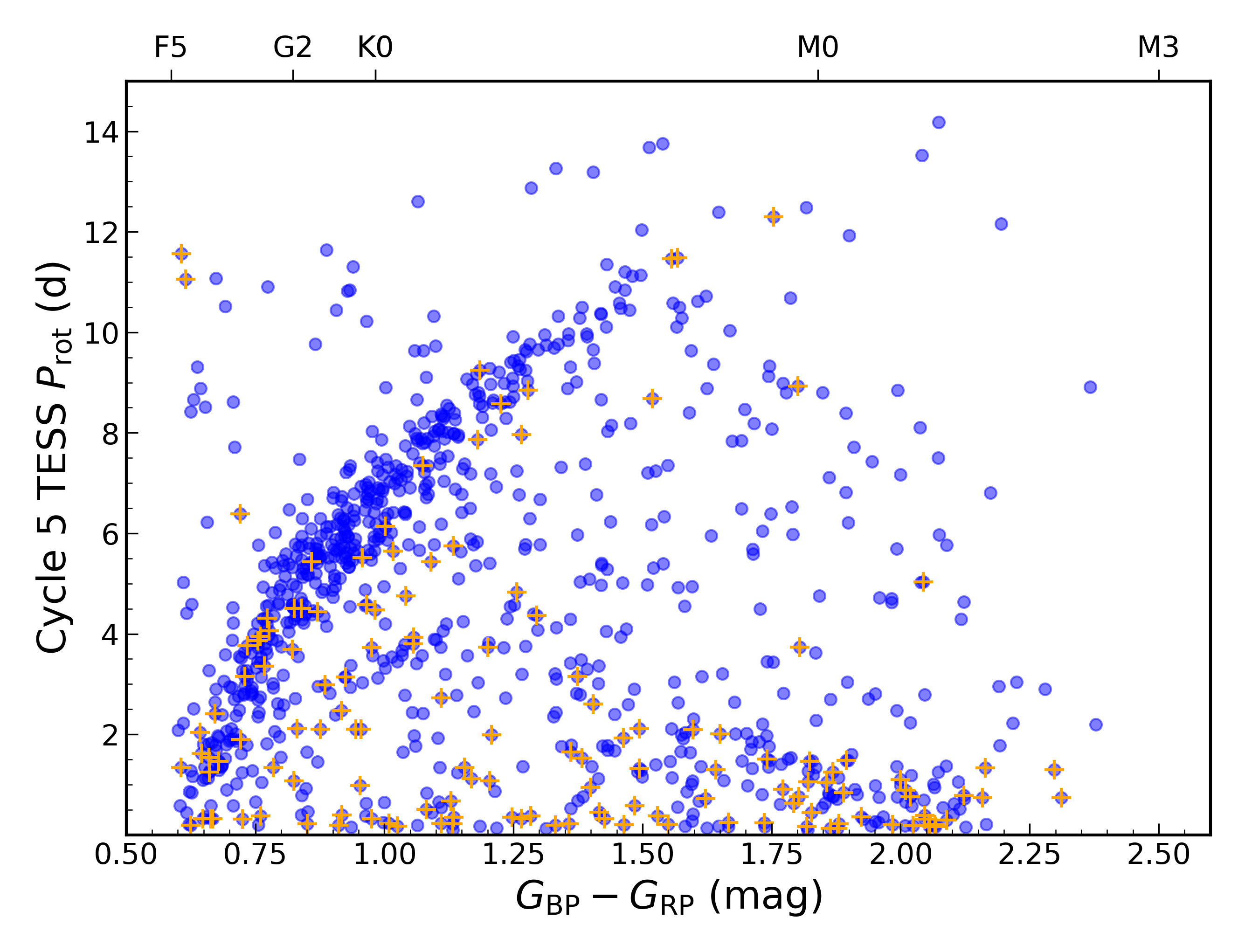}
\caption{TESS Cycle 5 CPD. All the periods shown here have powers above the threshold for this cycle. \prot\ measurements identified for removal due to source confusion either by GVC objects or by other cluster members are marked with orange crosses. A significant fraction of these \prot\ are $<$0.5 days, faster than we would expect for e.g., stars earlier than mid-K at this age \citep[see][]{fritz2021_periods}.
\label{fig:blending_cpd}}
\end{center}
\end{figure}

Figure~\ref{fig:blending_cpd} shows the impact of removing \prot\ contaminated either by GVC sources or by other cluster members on the \prot\ distribution for Cycle 5.

\subsection{Correcting \prot\ Measurements That Result From Half-Period Harmonics}\label{sec:har}

Based on the half-period sequences of the slow-rotating sequences that appear in the pre-analysis CPDs in the left column of Figure~\ref{fig:before_after_CPD}, and on cases where we measured a period $\approx$half that reported in \citet{fritz2021_periods} for a given target (as seen in Figure~\ref{fig:comp} and discussed in Section~\ref{sec:compare}), we identified and corrected \prot\ measurements resulting from half-period harmonics of the true period in our data. The availability of data from multiple TESS cycles was useful for identifying cases where we incorrectly assigned to a star half of its true \prot. Observations in Cycles 1, 3, and 5 were taken $\gapprox2$ years apart, long enough that the star spot distribution responsible for modulating our targets' light curves can evolve.\footnote{We do not expect the actual \prot\ to change on this timescale. The evidence from K2 data for the older Praesepe cluster is that while many light curves evolve on a very similar timescale, the derived \prot\ measurements are remarkably stable \citep{Rampalli2021}.} This may change the shape of the light curves, resolving confusion about which period is the real one and which is the harmonic.

Moreover, as shorter \prot\ are easier to detect in TESS data, in cases where we obtained a \prot\ for a star double that in one cycle relative to another, it is likely that the shorter \prot\ is not the true period. We searched for \prot\ $<$ 7.5 days that, when doubled, are $<$5\% different from a \prot\ measurement for the same star in another cycle. We flagged these shorter \prot\ as potential harmonics of the true periods.

Our two-term LS periodograms also helped with the identification of these cases. The two-term periodograms use an additional sinusoidal term when fitting the data, allowing for more robust \prot\ detection in some cases where the starspot distribution creates a signal that is more complex than can be represented by a single sine wave \citep[e.g., ``double dipping;''][]{Basri2018}. For each measurement, we compared the one-term and two-term LS periodograms. 

\begin{figure*}[!th]
%\centerline{\includegraphics[width=1.5\columnwidth, trim=0.65cm 0.6cm 0.6cm 0.cm, clip=True]{Periodogram_Comparison.png}}
\centerline{\includegraphics[width=1.5\columnwidth, trim=0.65cm 0.6cm 0.6cm 0.cm, clip=True]{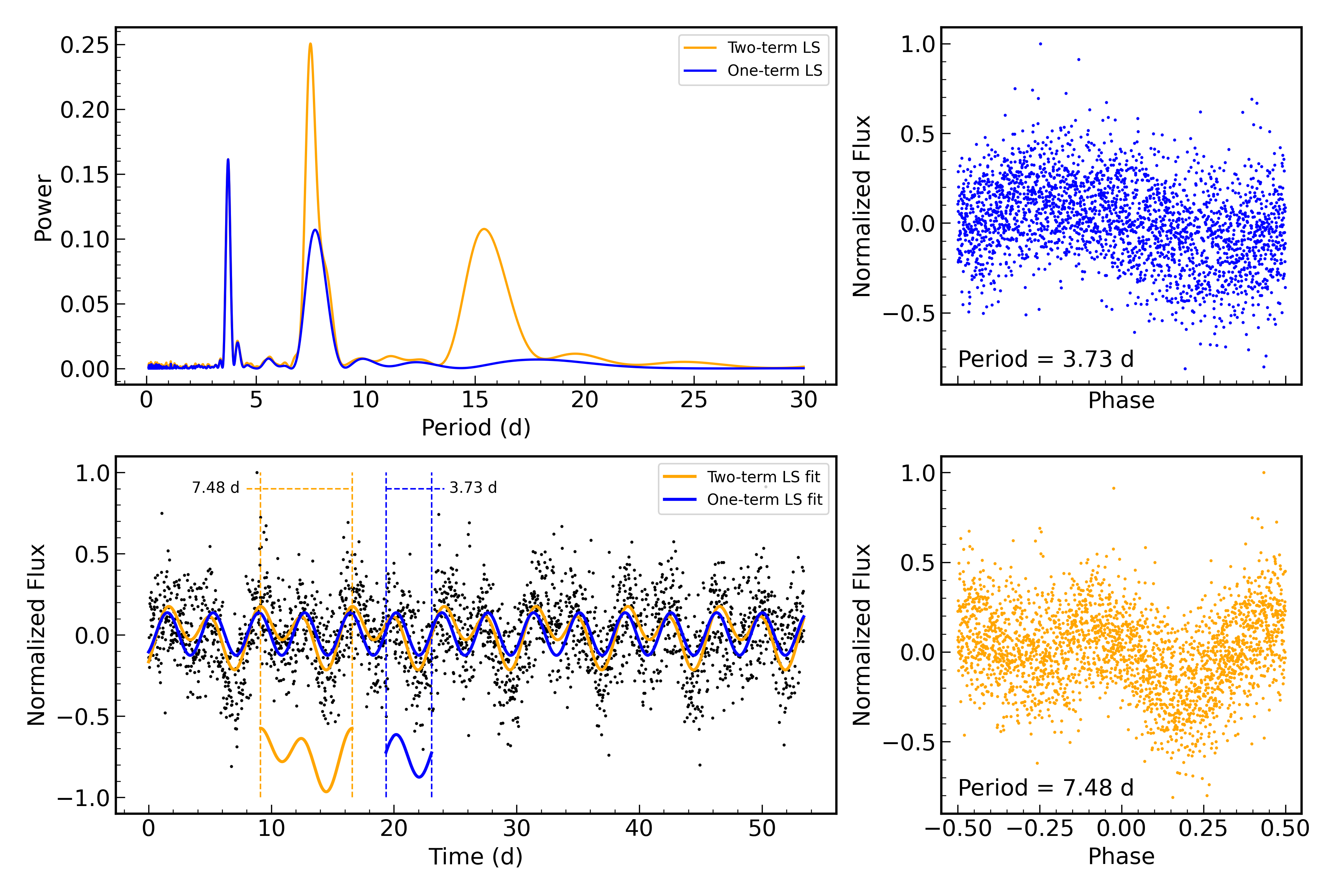}}
\caption{An example of a star for which using the two-term LS led to a period correction. The one- and two-term LS periodograms are shown at the top left. The corresponding best-fit LS models at the highest power period for each periodogram is plotted with the Cycle 5 light curve (bottom left). The location of the secondary peak in the one-term periodogram matches the location of the primary peak in the two-term periodogram (7.48~days). This corresponds to a \prot\ that is approximately double that measured using the location of the primary peak in the one-term periodogram (3.73~days). The original measurement at 3.73~days is a half-period harmonic of the true period and was corrected to 7.48~days. The top right panel shows the light curve phase-folded on the original \prot\ measurement, while the bottom right panel is the light curve phase-folded on the corrected measurement. The varying amplitude of the signal (i.e., the ``double dipping" causing the harmonic measurement) is evident in the bottom right, phase-folded light curve and is accurately modeled by the two-term periodogram.} 
\label{fig:pg_comp}
\end{figure*}

In cases where we measure the half-period harmonic of the true \prot\ for a target, the true \prot\ can appear as a lower-order peak in the one-term periodogram. We therefore flagged cases where the primary peak of the two-term periodogram matched the secondary peak of the one-term periodogram, provided that this secondary peak was at a period double that of the primary peak in the one-term periodogram. Figure~\ref{fig:pg_comp} shows a periodogram comparison that resulted in such a flag that helped provide evidence to correct a harmonic \prot\ measurement of a target for Cycle 5.

\begin{figure}
\begin{center}
\includegraphics[width=\columnwidth, trim=0.5cm 0.6cm 0.3cm 0.0cm, clip=True]{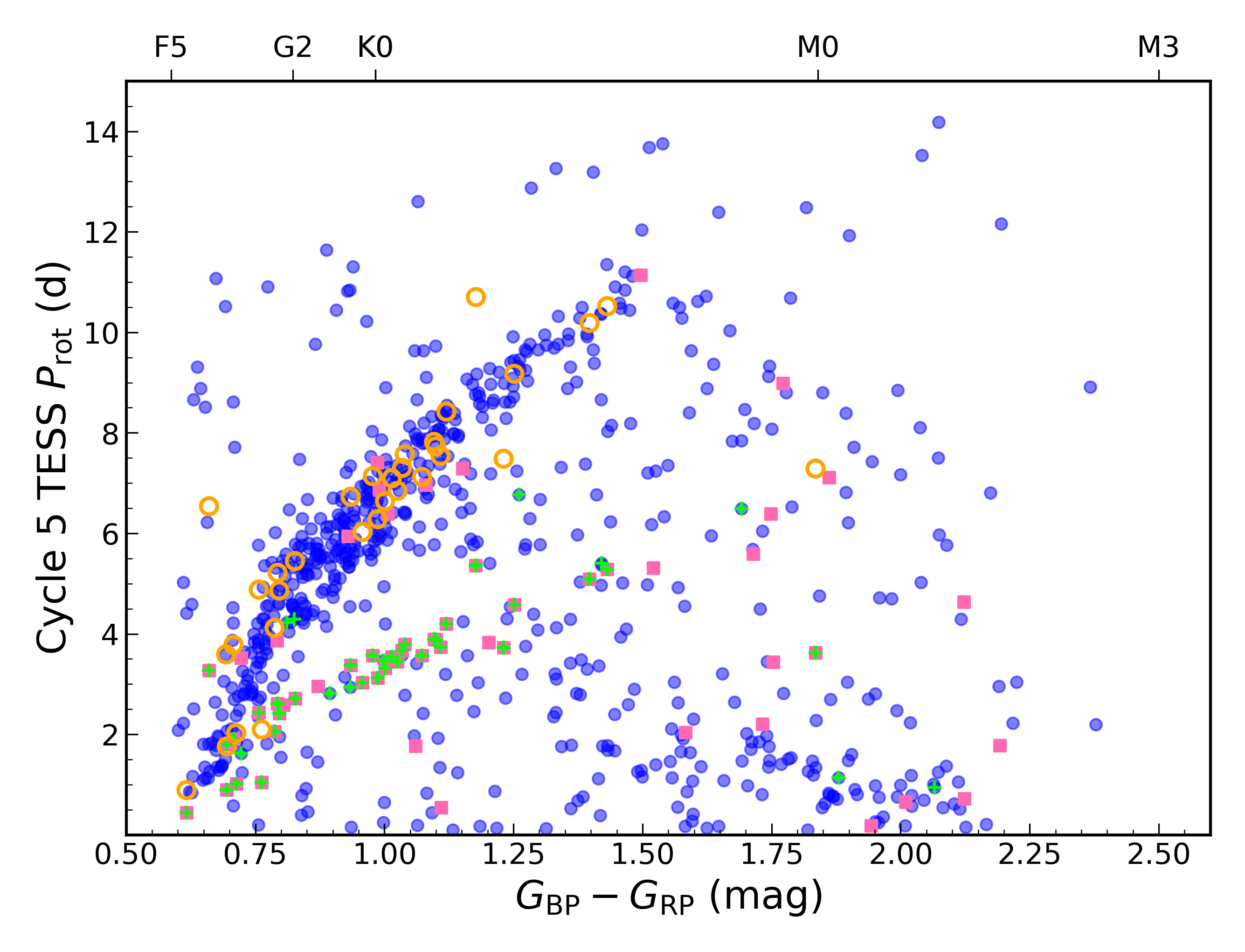}
\caption{TESS Cycle 5 CPD. Stars for which we measured twice the \prot\ measured for Cycle 5 (using the standard LS periodogram) in a different TESS cycle are marked with a green cross. \prot\ measurements identified as potential harmonic measurements using a comparison between one-term and two-term LS periodogram are  marked with pink squares. Where both these criteria are met for a given \prot\ measurement, we replaced our original measurement with the two-term LS measurement. The corrected measurements are marked with orange circles.
\label{fig:harmonic_cpd}}
\end{center}
\end{figure}

\begin{figure*}[!t]
%\centerline{\includegraphics[width=.79\textwidth, trim=0.6 0.6cm 0.6cm 0.6cm, clip=True]{Lit_Comp_Before_After.png}}
\centerline{\includegraphics[width=.79\textwidth, trim=0.6 0.6cm 0.6cm 0.6cm, clip=True]{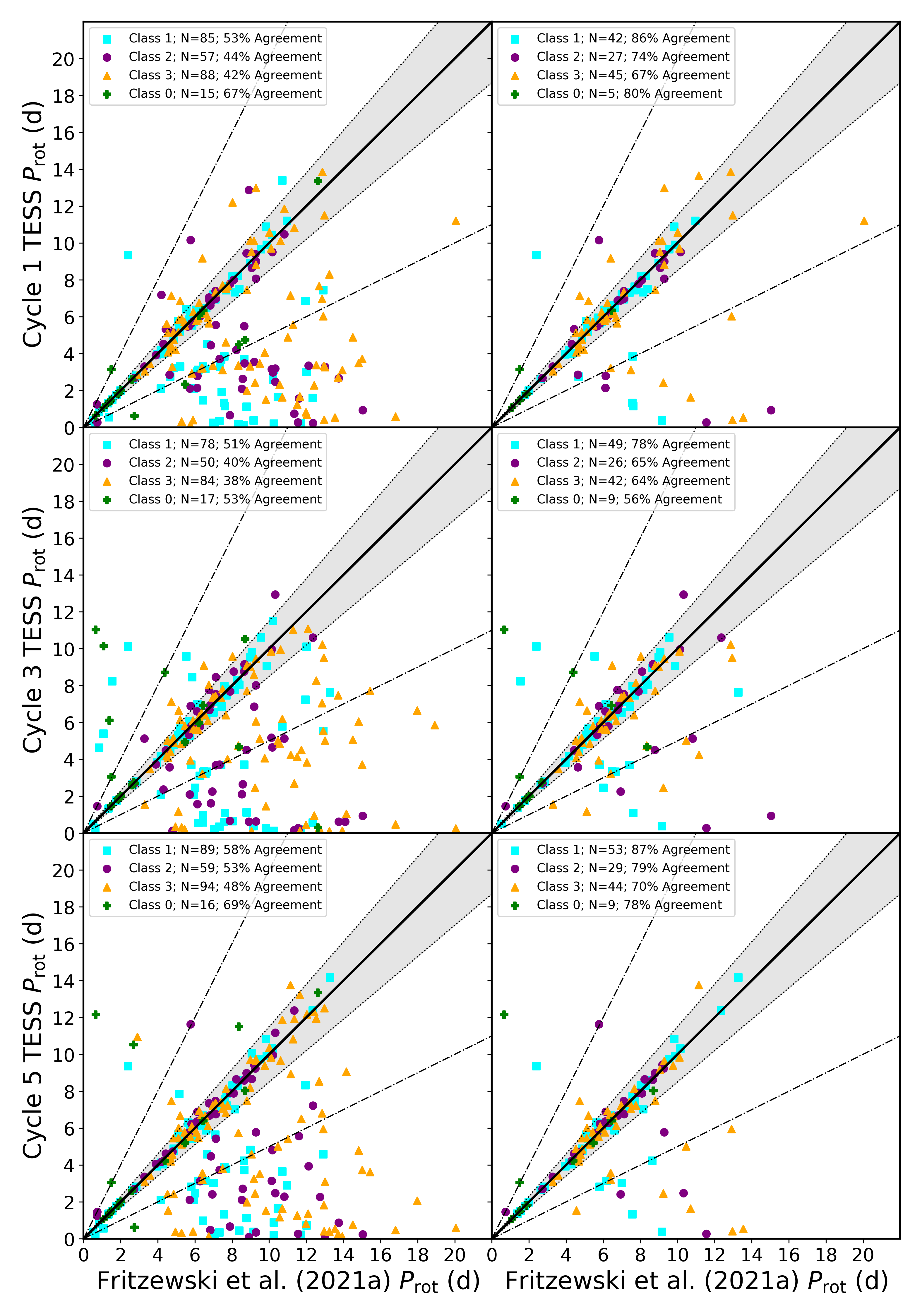}}
\caption{Comparisons of the \prot\ measured using TESS light curves from Cycles 1, 3, and 5 with those from \citet{fritz2021_periods} before (left) and after (right) our analysis removing our erroneous measurements. The shaded region indicates a difference of $\leq$15\% between the TESS and literature \prot\ values, which we take to be an agreement in the measurements. We also include 2:1 and 1:2 lines to indicate possible harmonic or subharmonic measurements. The classification assigned to each \prot\ measurement in \citet{fritz2021_periods} is indicated with a different symbol. Even before analyzing our results, we have a high \prot\ recovery rate (50--60\%) for the high-confidence (class 1) measurements from \citet{fritz2021_periods}, shown as filled cyan squares. Our initial agreement rate for lower-confidence measurements---especially those obtained using activity as a prior (class 3; filled orange triangles)---is systematically lower ($\approx$40--50\%). Our analysis of cases where our measurements differ significantly from those of \citet{fritz2021_periods} allows us to improve our recovery rate, so that it reaches $\gapprox80\%$ for the class 1 \prot, while other classes of \prot\ measurements saw slightly smaller improvements in agreement rate.}
\label{fig:comp}
\end{figure*}

For measurements that were flagged in both the periodogram comparison as well as the comparison with the measured \prot\ from other cycles, we replaced our original \prot\ measurement with that determined by the primary peak in the two-term LS periodogram. This resulted in the replacement of 15, 15, and 31 measurements for TESS Cycles 1, 3, and 5, respectively. In Figure~\ref{fig:harmonic_cpd}, we show the \prot\ we corrected in this manner in Cycle 5.

\subsection{Assessing the Impact of Applying Our Criteria}\label{sec:assess}

The cumulative effect of the steps described above to remove or correct spurious \prot\ measurements can be seen in the right column of Figure~\ref{fig:before_after_CPD}. Applying the periodogram power requirement removed by far the most measurements from the original CPD. Among the remaining high-power measurements, hundreds more were identified for removal due to source confusion, a necessary step for producing a \prot\ catalog for rich and relatively distant clusters like NGC~3532. 

Each CPD on the left side of Figure~\ref{fig:before_after_CPD} includes many spurious measurements with $\prot<1$ day that were removed in this step, as can be seen in the corresponding cleaned CPDs on the right. The changes in the \prot\ distribution due to the removal of measurements due to systematics, particularly in Cycle 1, is also  apparent in comparing the CPDs, as is the impact of our efforts to identify and correct half-period aliases. 

Most importantly, several hundred high-confidence \prot\ measurements were preserved in each cycle, as evidenced by the well-defined and well-populated sequences of slowly rotating stars in the CPDs in Figure~\ref{fig:before_after_CPD} and the remaining \prot\ measurements that agree with those from \citet{fritz2021_periods}. This gave us confidence that we could construct a robust, expanded rotational catalog for NGC~3532 based on these TESS data.

A summary of the number of \prot\ measurements removed by each step of the quality assessment of our measurements is shown in Table~\ref{table:removed}.

\begin{deluxetable}{lccc}
\centering 
\tabletypesize{\footnotesize}

\tablecaption{\prot\ measurements identified for removal\tablenotemark{a}
 \label{table:removed}}

\tablehead{
\colhead{} &
\colhead{Cycle 1} &
\colhead{Cycle 3} &
\colhead{Cycle 5}
}
\startdata
Targets observed & 1357 & 1054 & 1341 \\
\prot\ $>15$~d & 91 & 28 & 20 \\
Low power & 377 & 308 & 395 \\
Systematic error & 68 & 10 & N/A \\
Blending: GVC & 94 & 85 & 115 \\
Blending: Members & 69 & 58 & 82 \\
Unconfirmed rapid \prot\ & 15 & 15 & 31 \\
Remaining measurements & 643 & 550 & 698 \\
\enddata

\tablenotetext{a}{Rows 2--7 represent the number of \prot\ measurements identified for removal by each quality criterion after measuring \prot\ for each observation of each target. The numbers in these rows are those \prot\ measurements identified for removal that were not previously flagged for removal by any criterion represented by an above row.}

\vspace{-.2cm}
\end{deluxetable}

\section{Expanding the Rotational Census for NGC~3532} \label{sec:expand} 

\subsection{Revisiting the \citet{fritz2021_periods} \\ \prot\ Catalog}
\label{sec:compare}

We compared our automatically measured \prot\ prior to any quality cuts to those in the \citet{fritz2021_periods} catalog to assess their robustness. The results for each cycle are shown in the left column of Figure~\ref{fig:comp}, where we limit the comparison to stars for which we measured a \prot~$<$~15 days. 

About half of our \prot\ measurements from each TESS cycle differ by $<$15\% from those from \citet{fritz2021_periods}. The agreement is 48\% for Cycle 1 (117/245 periods), 44\% for Cycle 3 (101/229), and 54\% for Cycle 5 (139/258); our \prot\ measurements that agree with the literature appear as the red circles in Figure~\ref{fig:before_after_CPD}. When we combined the results for all three cycles, this overall recovery rate increased to 69\%---meaning that we recovered the \citet{fritz2021_periods} \prot\ 7/10 times in at least one of the TESS cycles.

Those agreement rates do not tell the full story, however: we were more likely to recover the photometric \prot\ measurements from \citet{fritz2021_periods}. The agreement rates between our measurements and these authors' highest-confidence photometric \prot\ (the class 1 periods) were 53\% (45/85), 51\% (40/78), and 58\% (52/89) for Cycles 1, 3, and 5 respectively. These agreement rates dropped to 44\% (25/57), 40\% (20/50), and 53\% (31/59) for the class 2 photometric \prot, and dropped again, to 42\% (37/88), 38\% (32/84), and 48\% (45/94), for the class 3 (activity-informed) \prot. Interestingly, the agreement rates for the class 0 \prot\ (potential aliases) was 67\% (10/15), 53\% (9/17), and 69\% (11/16) for Cycles 1, 3, and 5 respectively. 

The left column of Figure~\ref{fig:comp} suggests that where our measurements and those of \citet{fritz2021_periods} disagreed significantly, it is because our TESS measurements preferentially returned relatively short periods, as expected given the length of a sector. It also shows that we measured dozens of periods that fall on or near the 1:2 harmonic line, indicating again that harmonic \prot\ measurements appear in our sample and should be corrected where sufficient evidence is present. 

Following our analysis removing spurious measurements from our \prot\ sample, we retained measurements for 179 stars with existing \prot\ in the \citet{fritz2021_periods} catalog. We recovered the \prot\ from their catalog in at least one TESS cycle for 143 of these stars (80\%). The agreement rates between our measurements in each TESS cycle and the literature periods improved significantly following our analysis, as can be seen in the panels on the right side of Figure~\ref{fig:comp}. 

The post-analysis agreement rates between our measurements and these authors' class 1 periods improved to 86\% (36/42), 78\% (38/49), and 87\% (46/53) for Cycles 1, 3, and 5 respectively. The agreement rates for the class 2 photometric \prot\ improved to 79\% (20/27), 65\% (17/26), and 79\% (23/29), and for the class 3 (activity-informed) \prot, improved to 67\% (30/45), 64\% (27/42), and 70\% (31/44). The agreement rates for the class 0 \prot\ (potential aliases) also improved to 80\% (4/5), 56\% (5/9), and 78\% (7/9) for Cycles 1, 3, and 5, respectively.

\citet{Boyle2025} performed a comparison of TESS-derived \prot\ measured from CPM light curves and \prot\ from a K2 benchmark sample. After applying quality cuts to remove stars with significant photometric contamination and binaries, these authors found that $\gapprox$80\% of the TESS \prot\ $\leq$ 10 days agree with those obtained from K2 data, similar to our rate of agreement with the \citet{fritz2021_periods} class 1 \prot.

Many of the cases where we failed to recover the \prot\ from the \citet{fritz2021_periods} catalog are for slow rotators and/or for relatively faint stars, whose \prot\ are particularly difficult to obtain with TESS data. For the 36 stars for which we measured a \prot\ that we did not flag as problematic and that disagrees by more than 15\% with the \citet{fritz2021_periods} \prot, the median magnitude is $G=15.4$ mag.

Finally, there may be a small number of cases where we report an accurate \prot\ that does not agree with the literature. For example, there are  stars for which the \prot\ we measure is approximately twice that measured by \citet{fritz2021_periods}, as shown by the points falling on or near the 2:1 line in the panels of Figure~\ref{fig:comp}. It is possible that the spot distribution on the star at the time of observation caused these authors to measure the harmonic \prot.

\subsection{The new NGC~3532 \prot\ catalog}

Having identified likely inaccurate \prot\ measurements made using data from each TESS cycle, we consolidated the multiple \prot\ measurements of the same stars into a final set of \prot. Cases where the periods do not agree from cycle to cycle were not uncommon. For example, of the 179 stars from \citet{fritz2021_periods} for which we retained \prot\ measurements following our analysis, we measured a  \prot\ that agreed with the literature value for one TESS cycle and disagreed for another for 39 stars (22\%!).

Where \prot\ measurements obtained from multiple cycles survived our quality cuts, we calculated the median \prot\ and identified  which measurements agreed with this value within 10\%. For stars with multiple measurements that agreed with the median value, we recalculated the median using only those  measurements, and reported it as the final \prot\ for the target. If none of the measurements were within 10\% of the median \prot, we reported the longest \prot\ as the final \prot. We reasoned that this would generally select the true \prot, since rapid \prot\ are relatively easily detected and are more likely to be measured with multiple observations. For stars with only one \prot\ measurement after our quality cuts, we simply reported that \prot\ as the final \prot.

We classified our final \prot\ measurements according to the number of agreeing measurements used to arrive at the final measurement. Final \prot\ based on three agreeing measurements were our highest confidence \prot, and classified as Quality 3. Those based on two agreeing measurements are Quality 2, and those based on only one \prot\ measurement are classified as Quality 1.

\begin{figure*}[!th]
%\centerline{\includegraphics[width=2\columnwidth, trim=0.65cm 0.6cm 0.6cm 0.cm, clip=True]{Quality_CPDs.png}}
\centerline{\includegraphics[width=2\columnwidth, trim=0.65cm 0.6cm 0.6cm 0.cm, clip=True]{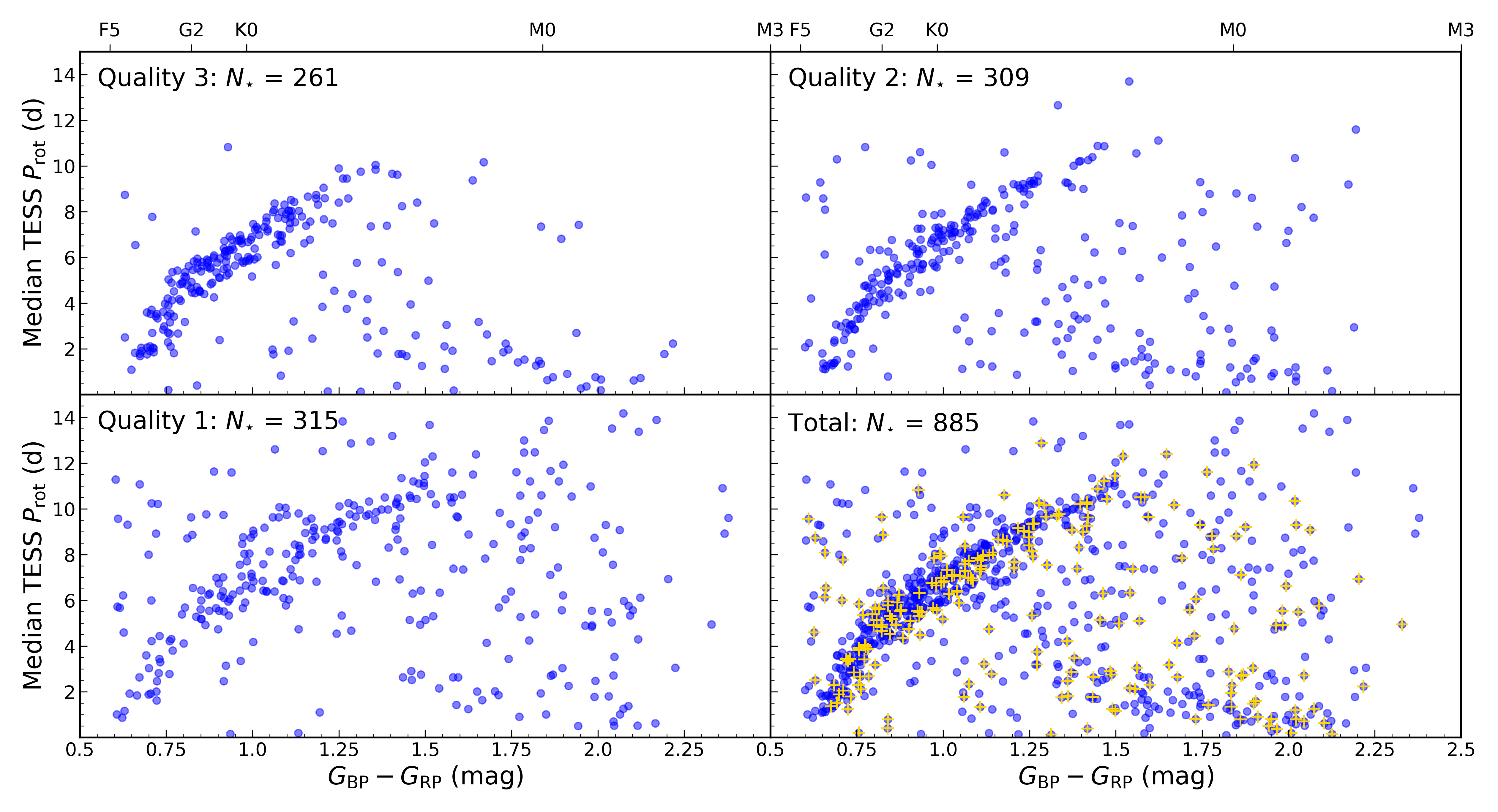}}
\caption{CPDs for NGC~3532 based on quality classification of the final \prot\ measurement reported for each star. Measurements classified as Quality 3 represent our highest confidence \prot, where the measured period was consistent across all three TESS cycles. Quality 2 \prot\ are measurements that agreed in two TESS cycles, and Quality 1 \prot\ were measured in only one cycle, sometimes corresponding to the longest \prot\ measured when measurements in multiple cycles disagreed. In total, we make new \prot\ measurements for 885 stars, all of which are plotted in the bottom right CPD. The yellow crosses in that CPD correspond to the  228 stars that have \texttt{HDBSCAN} membership probabilities of $<$50\% as indicated by \cite{Hunt2023} and for which we make new \prot\ measurements. Many of these lower-probability members fall on the slow-rotating sequence, suggesting that they are a part of the coeval NGC~3532 population and are likely cluster members.} 
\label{fig:quality_cpd}
\end{figure*}

In this manner, we created an expanded \prot\ catalog that includes 885 members of NGC~3532: 261 with Quality 3 \prot, 309 with Quality 2, and 315 with Quality 1. In Figure~\ref{fig:quality_cpd}, we show the CPDs for each quality group, as well as the CPD containing all the final \prot\ measurements. In that CPD (bottom right panel), we also highlight the 228 stars with \texttt{HDBSCAN} membership probabilities $<$50\% in \citet{Hunt2023}. Many of these stars have \prot\ measurements that fall on the slow-rotating sequence for NGC~3532, suggesting that they are part of the coeval cluster population. We are therefore able to confirm these at least as likely cluster members.

The CPD in the bottom right panel of Figure~\ref{fig:quality_cpd} containing all of our final \prot\ measurements features a well-populated slow-rotating sequence as well as a fast-rotating sequence that includes K stars still converging onto the slow sequence. This is consistent with the findings of \citet{fritz2021_periods} and expected for a cluster of the age of NGC~3532. Most of our new \prot\ measurements fall in regions of the CPD already populated with measurements from \citet{fritz2021_periods}. We added \prot\ measurements for mid-to-late F stars on the slow-rotating sequence for which there were previously none. The slow-rotating sequence defined by our \prot\ measurements does not extend to include M dwarfs although they are present in the \citet{fritz2021_periods} catalog. This is the result of the limitations of TESS in measuring \prot\ for slow-rotators and faint stars. Meanwhile, the relatively small population of fast-rotating G stars that fall beneath the slow-rotating sequences in the CPDs in Figure~\ref{fig:quality_cpd} are not likely to be accurate for a cluster of the age of NGC~3532 and might be the result of unidentified cases of photometric blending, or else are tidally-interacting binaries.\footnote{\citet{fritz2021_periods} provides a detailed discussion of the morphology of the CPD for NGC~3532 and comparisons to some existing semi-empirical models. \citet{gyrointerp2023} and \citet{ChronoFlow2025} place the period data for NGC~3532 in the context of many other open clusters and use those data to build new empirical gyrochronology models (\texttt{gyro-interp} and \texttt{ChronoFlow}, respectively).}

\begin{figure}[!t]
%\centerline{\includegraphics[width=\columnwidth, trim=0.65cm 0.6cm 0.6cm 0.cm, clip=True]{Lit_Comp_Final.png}}
\centerline{\includegraphics[width=\columnwidth, trim=0.65cm 0.6cm 0.6cm 0.cm, clip=True]{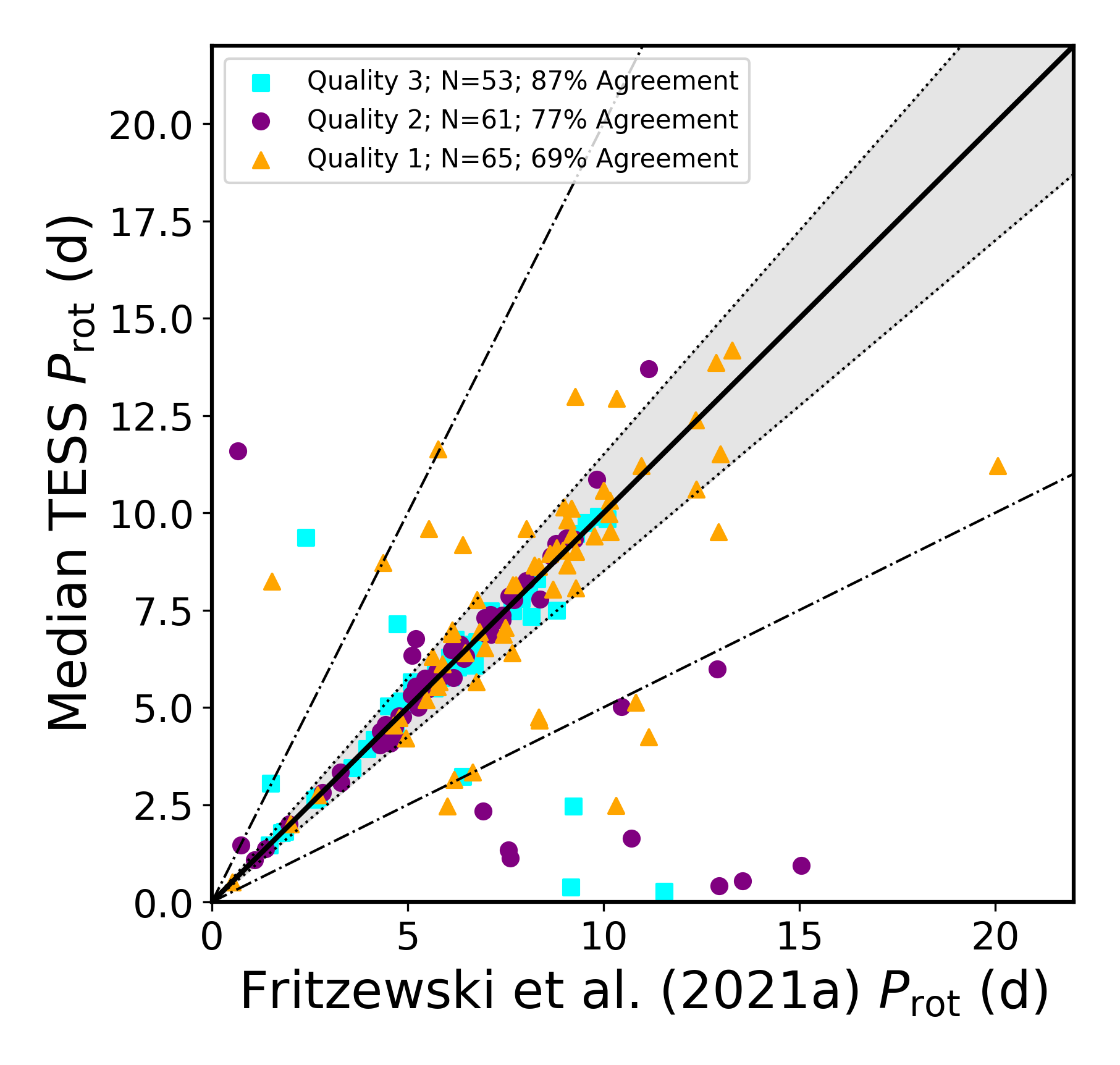}}
\caption{Comparison of our final \prot\ measurements with those from \citet{fritz2021_periods}. Different symbols are given to indicate the quality classification we assigned our \prot\ measurements. Our higher-confidence \prot\ measurements, which are  validated with observations from multiple TESS cycles (filled cyan squares, 3 cycles; filled purple circles, 2) have higher agreement rates with the measurements from \citet{fritz2021_periods}.
\label{fig:final_comp}}
\end{figure} 

We repeated our comparison with \prot\ from \citet{fritz2021_periods}, now using our final \prot\ measurements and comparing agreement rates for our quality classifications. The resulting \prot\ comparison plot is shown in Figure~\ref{fig:final_comp}. The overall agreement between the two sets of measurements is 77\%, but it does vary significantly according to our quality classifications: it is 87\% for our Quality 3 periods, and drops to 77\% and 69\% for our Quality 2 and 1 periods respectively.\footnote{Although we recovered \prot\ for 80\% of our remaining targets post-analysis with existing measurements from \citet{fritz2021_periods} in at least one TESS cycle, our final \prot\ selection process was not informed by existing \prot\ measurements. For five stars, the final \prot\ we selected does not agree with the literature \prot\ even though we measured \prot\ that agreed with the existing measurement for one TESS cycle.} These agreement rates serve as an estimate of an upper limit of the proportion of our \prot\ measurements that are accurate in each quality group. We are therefore more confident in \prot\ measurements confirmed with multiple observations and suspect that most of the remaining disagreement is caused by blended signals not recognized by our procedure.

\begin{figure*}[!th]
%\centerline{\includegraphics[width=2\columnwidth, trim=0.65cm 0.6cm 0.6cm 0.cm, clip=True]{TESS_Lit_CPD.png}}
\centerline{\includegraphics[width=2\columnwidth, trim=0.65cm 0.6cm 0.6cm 0.cm, clip=True]{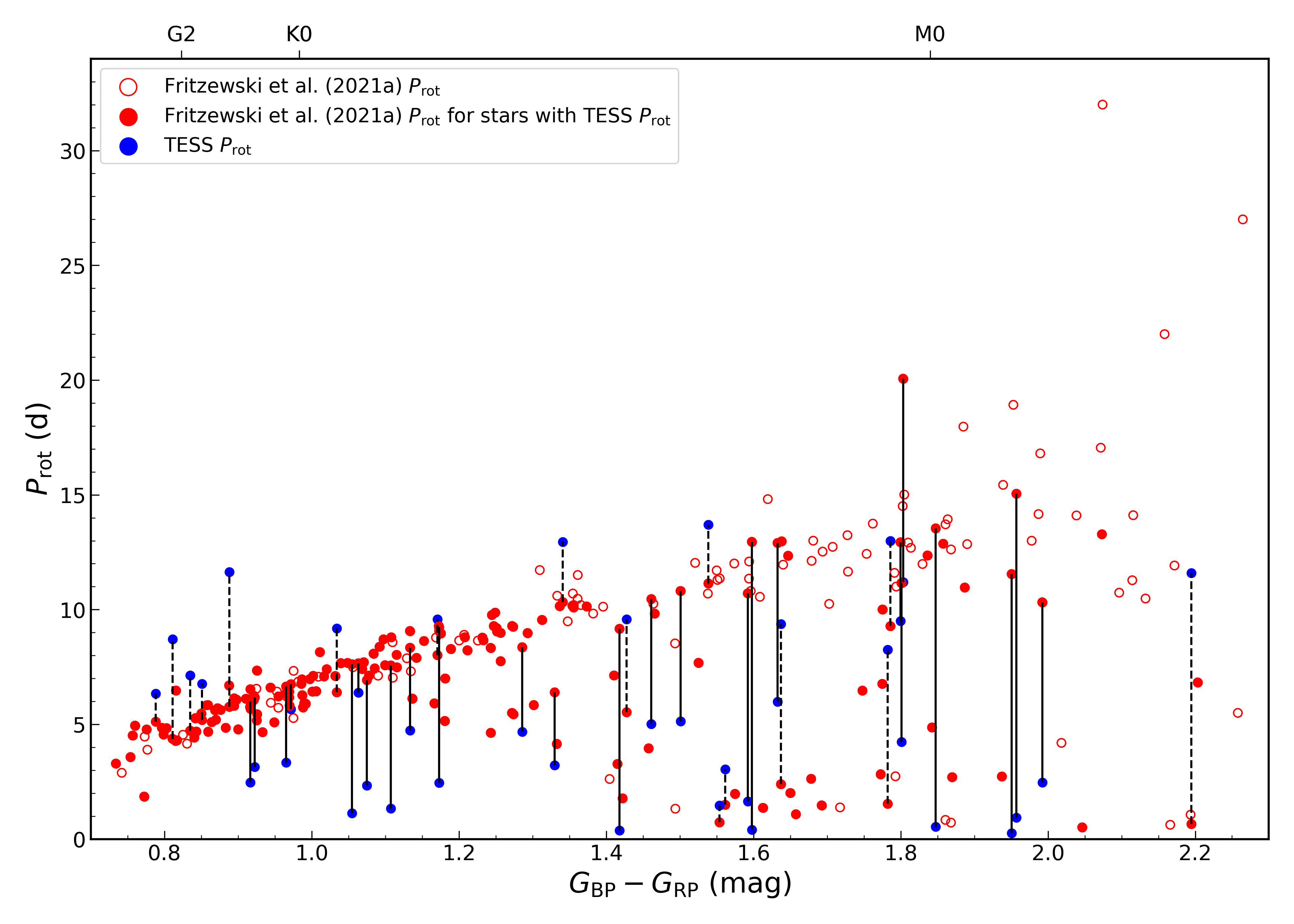}}
\caption{CPD for NGC~3532 comparing the \prot\ reported by \citet{fritz2021_periods} and our measurements. The filled red circles are literature \prot\ for stars for which we obtained a \prot, while the open red circles are literature \prot\ for stars for which we did not. The filled blue circles are our 41 newly measured \prot\ that disagreed by $>$15\% with the \citet{fritz2021_periods} \prot.  Black lines connect our measurements and the \citet{fritz2021_periods} periods when our \prot\ is shorter. Dashed lines are used for the opposite case. In most cases where the \prot\ measurements disagree significantly, we measured a \prot\ that is much more rapid than that in \citet{fritz2021_periods}, suggesting that a fraction of our measurements remain impacted by blending.} 
\label{fig:cpd_comp}
\end{figure*}

In Figure~\ref{fig:cpd_comp}, we show a CPD using \prot\ from \citet{fritz2021_periods} and highlight the 41 cases where our final measurement disagrees by $>$15\% with the literature \prot\ for the same star. Generally, we are more likely to measure relatively rapid \prot\ due to their systematically higher powers in LS periodograms. Some erroneous rapid \prot\ measurements persist in our final catalog, most likely due to unidentified cases of photometric blending. Our recovery of \prot\ measurements for slower rotators is comparatively low presumably due to a combination of this contamination, the fact that slower rotators tend to be relatively faint, and that, as mentioned, TESS has limited ability to detect variability on longer time scales. Figure~\ref{fig:final_comp} also suggests that we may not have successfully identified all of the harmonic measurements we made since some inaccurate measurements  are $\approx$half the \prot\ from \citet{fritz2021_periods}.

Our recovery rate of \citet{fritz2021_periods} \prot\ is highest for these authors' highest confidence measurements, those directly measured from photometry. Similarly, the agreement rate between our \prot\ and those from \citet{fritz2021_periods} is highest for our highest confidence (Quality 3) measurements at 87\%. We conclude that this justifies using \prot\ measured from TESS observations to build \prot\ catalogs for NGC~3532 and other crowded open clusters. However, this must be done with appropriate caution to identify inaccurate \prot\ measurements, especially for fainter stars and slower rotators, and ideally using multiple TESS observations to validate \prot\ measurements.

While our recovery rate is lower for lower-confidence \citet{fritz2021_periods} \prot\ measurements, we did recover a substantial fraction of the \prot\ these authors measured using predictions based on a target's chromospheric activity, underscoring the usefulness of this approach. In general, however, many of these lower-confidence \prot\ are for stars for which it is difficult to measure \prot\ with TESS. And our efforts to produce our expanded \prot\ catalog for NGC~3532 are successful in identifying most of our inaccurate measurements for these targets.

In Table~\ref{tbl:catalog}, we provide a description of the columns in our final \prot\ catalog for NGC~3532. 

\begin{deluxetable}{@{}ll@{}}
\digitalasset
\tabletypesize{\footnotesize} 

\tablecaption{Columns in Our NGC~3532 Rotation Catalog \label{tbl:catalog}}

\tablehead{
\colhead{Column} & \colhead{Description}
}

\startdata
1      & Gaia DR3 source designation \\
2      & HDBSCAN membership probability \citep{Hunt2023} \\
3, 4   & R.A., decl. at Epoch = 2016.0 \\
5, 6   & Parallax and 1$\sigma$ uncertainty \\
7   & Gaia DR3 $G$ magnitude \\
8     & Gaia DR3 color \gbr\ \\
9     & Gaia DR3 Re-normalized Unit Weight Error \\
10     & Binary flag (RUWE$>1.4$)\tablenotemark{a} \\
11     & TESS-based rotation period \prot \\
12     & \prot\ quality classification\tablenotemark{b} \\
13, 14     & \prot\ and periodogram power (Cycle 1)\tablenotemark{c} \\
15, 16     & \prot\ and periodogram power (Cycle 3)\tablenotemark{c} \\
17, 18     & \prot\ and periodogram power (Cycle 5)\tablenotemark{c} \\
19, 20, 21     & Low power flag (Cycles 1, 3, and 5)\tablenotemark{a} \\
22, 23, 24     & Source confusion flag (Cycles 1, 3, and 5) \tablenotemark{a} \\
25, 26     & Systematic error flag (Cycles 1 and 3)\tablenotemark{a} \\
27, 28, 29     & Corrected harmonic period flag (Cycles 1, 3, and 5)\tablenotemark{a} \\
30, 31, 32     & Vanished rapid period flag (Cycles 1, 3, and 5)\tablenotemark{a} \\
33     & Literature \prot\ \citep{fritz2021_periods}\tablenotemark{d} \\
34     & Literature \prot\ class \citep{fritz2021_periods}\tablenotemark{d} \\
\enddata
\tablecomments{The contents of this table are published as a machine-readable table in the electronic edition of the Journal.}
\tablenotetext{a}{Columns for data flags use zero to represent no flag and one to represent a flag.}
\tablenotetext{b}{The quality classification for a given TESS rotation period is equal to the number of TESS observations used to confirm that rotation period measurement.}
\tablenotetext{c}{Columns for rotation periods and periodogram powers for individual TESS cycles use zero to represent no available data and negative values to represent values flagged for removal.}
\tablenotetext{d}{Columns with data from the existing rotation period catalog for NGC 3532 \citep{fritz2021_periods} use $-1$ to represent no available data.}

\vspace{-0.25in}

\end{deluxetable}

\section{Conclusion} \label{sec:con}

TESS provides us with the ability to measure \prot\ for low-mass stars across most of the sky.  However, TESS's large pixels mean that blending and source confusion are a concern when constructing \prot\ catalogs for relatively distant and/or crowded fields. 

We used the relatively young, relatively distant, and rich open cluster NGC~3532 as a test case for  a procedure for removing inaccurate \prot\ measurements from an automatically generated period catalog produced from TESS data. While we compared our results to the existing \cite{fritz2021_periods} \prot\ catalog for the cluster to test our approach's usefulness and reliability, ultimately we relied on data from TESS and Gaia to evaluate our periods, and our approach can therefore be applied to clusters without existing rotation catalogs. 

The primary source of inaccurate \prot\ measurements that seem to be of good quality is the presence of neighboring variable sources with comparatively strong signals that cause photometric blending in target light curves. By identifying stars with such neighbors and removing our measurements of their \prot\ from our rotation catalog, we were able to produce a reliable, expanded period catalog for NGC~3532 that more than triples the size of the sample for this cluster: \citet{fritz2021_periods} published \prot\ measurements for 279 stars, and our catalog includes 885 stars, 706 of which did not previously have a \prot\ measurement. 

We assessed our approach by comparing our \prot\ measurements to those obtained by \citet{fritz2021_periods} several times. The first was after running a standard Lomb-Scargle-based search for periods in all of the TESS data available for the \cite{Hunt2023} cluster members. We found that we could recover accurate (within 15\%) \prot\ for 69\% of the \citet{fritz2021_periods} stars in at least one cycle of TESS data before any analysis of the quality of our measurements. Post this analysis, the overall agreement rate between our measurements increased to 77\%, but it is 86\% for the highest quality \citet{fritz2021_periods} \prot\ measurements and 87\% for our own highest-confidence measurements, a remarkable result given the apparent challenges a crowded field like that of NGC~3532 poses to TESS.

In our final catalog, there were 41 stars for which our measured \prot\ disagreed with the \prot\ of \citet{fritz2021_periods} by more than 15\%. For each of these stars, we inspected the light curves and periodograms produced from the targets' pixels and surrounding pixels to understand what might be causing inaccurate measurements that persist despite our attempts to filter them out. In many of these cases, there was a signal from a neighboring source that interfered with the measurement for our target. In some cases, this led to a measurement of variability that is clearly accurate for a nearby source and not for our target. In other cases, the photometric contamination is sufficiently complicated that it is not clear if the period reported is accurate for any nearby source. As discussed in Section~\ref{sec:assess}, we removed from our catalog measurements caused by source confusion only if we were able to identify the true source of the detected variability. This prevented us from identifying every case of photometric contamination in our catalog. There also may remain a few cases where we measured the harmonic of the true \prot\ and did not find enough evidence to correct this measurement in our analysis.  This suggests that inaccurate \prot\ measurements remain in our expanded catalog, particularly for (relatively) slow rotators and faint stars. 

We benefited from having three cycles' worth of TESS data for  NGC~3532, which gave us three separate opportunities to measure \prot\ for many of our targets. Interestingly, our analysis of the data from the three cycles did not always return the same \prot\ for our targets. The changing orientation of the telescope between cycles causes different pixel geometries relative to the field, which can impact our measurements. There were also clear systematics present in one or the other of the cycles. As a result, for some stars it was possible to measure an accurate \prot\ for a target using FFIs from one TESS cycle, but not another. We used the agreement in our measurements between cycles to assign a quality to our \prot.

In some cases, we removed from our sample \prot\ measurements that agreed with the existing measurement from \citet{fritz2021_periods} for a given star. This occurred most frequently due to our largest quality cut: the periodogram power threshold. Across the three studied TESS cycles, we removed 209 \prot\ measurements for stars with existing measurements from the literature catalog with this quality cut. Of these measurements, 50 agreed with that from the literature within 15\%. While 3/4 of the measurements we removed with this quality cut are spurious, it may be possible to recover more accurate low-power measurements, especially using additional observations as TESS continues to survey the sky.

NGC~3532 was recently observed in TESS Cycle 7 and will be observed again in Cycles 8 and 9. In Cycle 9, NGC~3532 will be observed in three consecutive sectors, possibly enabling measurement of higher-confidence \prot\ for slow rotators. A re-analysis of our catalog after these observations should result in a refinement and possibly an expansion of our period catalog. As is, the catalog presented here is one of the largest assembled for an open cluster, and should contribute to efforts to calibrate gyrochronology for low-mass main-sequence stars. 

Further development and application of the strategies introduced here will allow for more reliable measurement of \prot\ in distant, crowded clusters that continue to be observed by TESS. This opens the possibility of producing expansive \prot\ catalogs for open clusters (or other co-moving, coeval structures) at  unstudied or understudied ages and metallicities, thereby contributing new benchmarks to studies of gyrochronology.

%%%%%%%%%%%%%%%%%%%%%%%%%%%%%%%%%%%%%%%%%%%%%%%%%%%%%%%%%%%%%%%%%%%%%%%%%%%%%%%%%%%%%%%%%%%%%%%%%%%%%%%%%%%%%%%%%%
%% IMPORTANT! The old "\acknowledgment" command has be depreciated. It was
%% not robust enough to handle our new dual anonymous review requirements and
%% thus been replaced with the acknowledgment environment. If you try to 
%% compile with \acknowledgment you will get an error print to the screen
%% and in the compiled pdf.

\begin{acknowledgments}

We thank the anonymous reviewer for their careful reading and comments that improved the manuscript.

% Funding
This work was supported by NSF Astronomy \& Astrophysics Grant AST-2009840 and NASA TESS GI grants No. 80NSSC22K0299 (program ID G04217), 80NSSC23K0369 (program ID G05157), 80NSSC23K0370 (program ID G05028), and 80NSSC24K0501 (program ID G06163). M.A.A.~acknowledges support from a Fulbright U.S.~Scholar grant co-funded by the Nouvelle-Aquitaine Regional Council and the Franco-American Fulbright Commission. M.A.A.~also acknowledges support from a Chrétien International Research Grant from the American Astronomical Society.

% TESS, MAST
This paper includes data collected by the TESS mission, which are publicly available from the Mikulski Archive for Space Telescopes (MAST). Funding for the TESS mission is provided by NASA’s Science Mission directorate. The TESS data used in this paper can be found in MAST: \dataset[10.17909/0cp4-2j79]{http://dx.doi.org/10.17909/0cp4-2j79}.

% Gaia
This work has made use of data from the European Space Agency (ESA)
mission Gaia,\footnote{\url{https://www.cosmos.esa.int/gaia}} processed by
the Gaia Data Processing and Analysis Consortium (DPAC).\footnote{\url{https://www.cosmos.esa.int/web/gaia/dpac/consortium}} Funding
for the DPAC has been provided by national institutions, in particular
the institutions participating in the Gaia Multilateral Agreement.
This research also made use of public auxiliary data provided by ESA/Gaia/DPAC/CU5 and prepared by Carine Babusiaux.

The Digitized Sky Surveys were produced at the Space Telescope Science Institute under U.S. Government grant NAG W-2166. The images of these surveys are based on photographic data obtained using the Oschin Schmidt Telescope on Palomar Mountain and the UK Schmidt Telescope. The plates were processed into the present compressed digital form with the permission of these institutions.

% SIMBAD, Vizier, ADS
This research has also made use of NASA's Astrophysics Data System, 
and the VizieR \citep{vizier} and SIMBAD \citep{simbad} databases 
operated at CDS, Strasbourg, France.

\end{acknowledgments}

%% To help institutions obtain information on the effectiveness of their 
%% telescopes the AAS Journals has created a group of keywords for telescope 
%% facilities.
%
%% Following the acknowledgments section, use the following syntax and the
%% \facility{} or \facilities{} macros to list the keywords of facilities used 
%% in the research for the paper.  Each keyword is check against the master 
%% list during copy editing.  Individual instruments can be provided in 
%% parentheses, after the keyword, but they are not verified.

\vspace{5mm}
\facilities{TESS, Gaia}

%% Similar to \facility{}, there is the optional \software command to allow 
%% authors a place to specify which programs were used during the creation of 
%% the manuscript. Authors should list each code and include either a
%% citation or url to the code inside ()s when available.

\software{  astropy \citep{Astropy2013, Astropy2018},
            astroquery \citep{astroquery}, 
            Matplotlib \citep{matplotlib}, 
            NumPy \citep{Numpy2020}, 
            SciPy \citep{scipy}, 
            TESScut \citep{brasseur2019}, 
            unpopular \citep{Hattori2022},
            lightkurve
            \citep{lightkurve}
        }

%% Appendix material should be preceded with a single \appendix command.
%% There should be a \section command for each appendix. Mark appendix
%% subsections with the same markup you use in the main body of the paper.

%% Each Appendix (indicated with \section) will be lettered A, B, C, etc.
%% The equation counter will reset when it encounters the \appendix
%% command and will number appendix equations (A1), (A2), etc. The
%% Figure and Table counter will not reset.

%%%%%%%%%%%%%%%%%%%%%%%%%%%%%%%%%%%%%%%%%%%%%%%%%%%%%%%%%%%%%%%%%%%%%%%%%%%%%%%%%%%%%%%%%%%%%%%%%%%%%%%%%%%%%%%%%%

%% For this sample we use BibTeX plus aasjournals.bst to generate the
%% the bibliography. The sample631.bib file was populated from ADS. To
%% get the citations to show in the compiled file do the following:
%%
%% pdflatex sample631.tex
%% bibtext sample631
%% pdflatex sample631.tex
%% pdflatex sample631.tex
%\clearpage

%\bibliography{refs.bib}{}

\begin{thebibliography}{}
\expandafter\ifx\csname natexlab\endcsname\relax\def\natexlab#1{#1}\fi
\providecommand{\url}[1]{\href{#1}{#1}}
\providecommand{\dodoi}[1]{doi:~\href{http://doi.org/#1}{\nolinkurl{#1}}}
\providecommand{\doeprint}[1]{\href{http://ascl.net/#1}{\nolinkurl{http://ascl.net/#1}}}
\providecommand{\doarXiv}[1]{\href{https://arxiv.org/abs/#1}{\nolinkurl{https://arxiv.org/abs/#1}}}

\bibitem[{M.~A. {Ag{\"u}eros} {et~al.}(2025){Ag{\"u}eros}, {Curtis}, {N{\'u}{\~n}ez}, {Burhenne}, {Rothstein}, {Shaham}, {Singh}, {Bergeron}, {Kilic}, {Covey}, \& {Douglas}}]{agueros2025}
{Ag{\"u}eros}, M.~A., {Curtis}, J.~L., {N{\'u}{\~n}ez}, A., {et~al.} 2025, \bibinfo{title}{{Crowning the Queen: Membership, Age, Rotation, and Activity for the Open Cluster Coma Berenices},} \apj, 993, 144, \dodoi{10.3847/1538-4357/ae03a3}

\bibitem[{F. {Anthony} {et~al.}(2022){Anthony}, {N{\'u}{\~n}ez}, {Ag{\"u}eros}, {Curtis}, {do Nascimento}, {Machado}, {Mann}, {Newton}, {Rampalli}, {Thao}, \& {Wood}}]{anthony2022}
{Anthony}, F., {N{\'u}{\~n}ez}, A., {Ag{\"u}eros}, M.~A., {et~al.} 2022, \bibinfo{title}{{Activity and Rotation of Nearby Field M Dwarfs in the TESS Southern Continuous Viewing Zone},} \aj, 163, 257, \dodoi{10.3847/1538-3881/ac6110}

\bibitem[{ {Astropy Collaboration} {et~al.}(2013){Astropy Collaboration}, {Robitaille}, {Tollerud}, {Greenfield}, {Droettboom}, {Bray}, {Aldcroft}, {Davis}, {Ginsburg}, {Price-Whelan}, {Kerzendorf}, {Conley}, {Crighton}, {Barbary}, {Muna}, {Ferguson}, {Grollier}, {Parikh}, {Nair}, {Unther}, {Deil}, {Woillez}, {Conseil}, {Kramer}, {Turner}, {Singer}, {Fox}, {Weaver}, {Zabalza}, {Edwards}, {Azalee Bostroem}, {Burke}, {Casey}, {Crawford}, {Dencheva}, {Ely}, {Jenness}, {Labrie}, {Lim}, {Pierfederici}, {Pontzen}, {Ptak}, {Refsdal}, {Servillat}, \& {Streicher}}]{Astropy2013}
{Astropy Collaboration}, {Robitaille}, T.~P., {Tollerud}, E.~J., {et~al.} 2013, \bibinfo{title}{{Astropy: A community Python package for astronomy},} \aap, 558, A33, \dodoi{10.1051/0004-6361/201322068}

\bibitem[{S.~A. {Barnes}(2003){Barnes}}]{Barnes2003}
{Barnes}, S.~A. 2003, \bibinfo{title}{{On the Rotational Evolution of Solar- and Late-Type Stars, Its Magnetic Origins, and the Possibility of Stellar Gyrochronology},} \apj, 586, 464, \dodoi{10.1086/367639}

\bibitem[{S.~A. {Barnes}(2007){Barnes}}]{Barnes2007}
{Barnes}, S.~A. 2007, \bibinfo{title}{{Ages for Illustrative Field Stars Using Gyrochronology: Viability, Limitations, and Errors},} \apj, 669, 1167, \dodoi{10.1086/519295}

\bibitem[{G. {Basri} \& H.~T. {Nguyen}(2018){Basri} \& {Nguyen}}]{Basri2018}
{Basri}, G., \& {Nguyen}, H.~T. 2018, \bibinfo{title}{{Double Dipping: A New Relation between Stellar Rotation and Starspot Activity},} \apj, 863, 190, \dodoi{10.3847/1538-4357/aad3b6}

\bibitem[{W.~J. {Borucki} {et~al.}(2010){Borucki}, {Koch}, {Basri}, {Batalha}, {Brown}, {Caldwell}, {Caldwell}, {Christensen-Dalsgaard}, {Cochran}, {DeVore}, {Dunham}, {Dupree}, {Gautier}, {Geary}, {Gilliland}, {Gould}, {Howell}, {Jenkins}, {Kondo}, {Latham}, {Marcy}, {Meibom}, {Kjeldsen}, {Lissauer}, {Monet}, {Morrison}, {Sasselov}, {Tarter}, {Boss}, {Brownlee}, {Owen}, {Buzasi}, {Charbonneau}, {Doyle}, {Fortney}, {Ford}, {Holman}, {Seager}, {Steffen}, {Welsh}, {Rowe}, {Anderson}, {Buchhave}, {Ciardi}, {Walkowicz}, {Sherry}, {Horch}, {Isaacson}, {Everett}, {Fischer}, {Torres}, {Johnson}, {Endl}, {MacQueen}, {Bryson}, {Dotson}, {Haas}, {Kolodziejczak}, {Van Cleve}, {Chandrasekaran}, {Twicken}, {Quintana}, {Clarke}, {Allen}, {Li}, {Wu}, {Tenenbaum}, {Verner}, {Bruhweiler}, {Barnes}, \& {Prsa}}]{borucki2010}
{Borucki}, W.~J., {Koch}, D., {Basri}, G., {et~al.} 2010, \bibinfo{title}{{Kepler Planet-Detection Mission: Introduction and First Results},} Science, 327, 977, \dodoi{10.1126/science.1185402}

\bibitem[{L.~G. {Bouma} {et~al.}(2019){Bouma}, {Hartman}, {Bhatti}, {Winn}, \& {Bakos}}]{CDIPS}
{Bouma}, L.~G., {Hartman}, J.~D., {Bhatti}, W., {Winn}, J.~N., \& {Bakos}, G.~{\'A}. 2019, \bibinfo{title}{{Cluster Difference Imaging Photometric Survey. I. Light Curves of Stars in Open Clusters from TESS Sectors 6 and 7},} \apjs, 245, 13, \dodoi{10.3847/1538-4365/ab4a7e}

\bibitem[{L.~G. {Bouma} {et~al.}(2023){Bouma}, {Palumbo}, \& {Hillenbrand}}]{gyrointerp2023}
{Bouma}, L.~G., {Palumbo}, E.~K., \& {Hillenbrand}, L.~A. 2023, \bibinfo{title}{{The Empirical Limits of Gyrochronology},} \apjl, 947, L3, \dodoi{10.3847/2041-8213/acc589}

\bibitem[{A.~W. {Boyle} {et~al.}(2025){Boyle}, {Mann}, \& {Bush}}]{Boyle2025}
{Boyle}, A.~W., {Mann}, A.~W., \& {Bush}, J. 2025, \bibinfo{title}{{Quantifying the Limits of TESS Stellar Rotation Measurements with the K2-TESS Overlap},} \apj, 985, 233, \dodoi{10.3847/1538-4357/adcecc}

\bibitem[{C.~E. {Brasseur} {et~al.}(2019){Brasseur}, {Phillip}, {Fleming}, {Mullally}, \& {White}}]{brasseur2019}
{Brasseur}, C.~E., {Phillip}, C., {Fleming}, S.~W., {Mullally}, S.~E., \& {White}, R.~L. 2019, \bibinfo{title}{{Astrocut: Tools for creating cutouts of TESS images},}, Astrophysics Source Code Library, record ascl:1905.007 \doeprint{1905.007}

\bibitem[{R.~J. G.~B. Campello {et~al.}(2013)Campello, Moulavi, \& Sander}]{Campello2013}
Campello, R. J. G.~B., Moulavi, D., \& Sander, J. 2013, in Advances in Knowledge Discovery and Data Mining, ed. J.~Pei, V.~S. Tseng, L.~Cao, H.~Motoda, \& G.~Xu (Berlin, Heidelberg: Springer Berlin Heidelberg), 160--172

\bibitem[{T. {Cantat-Gaudin} {et~al.}(2018){Cantat-Gaudin}, {Jordi}, {Vallenari}, {Bragaglia}, {Balaguer-N{\'u}{\~n}ez}, {Soubiran}, {Bossini}, {Moitinho}, {Castro-Ginard}, {Krone-Martins}, {Casamiquela}, {Sordo}, \& {Carrera}}]{CG2018}
{Cantat-Gaudin}, T., {Jordi}, C., {Vallenari}, A., {et~al.} 2018, \bibinfo{title}{{A Gaia DR2 view of the open cluster population in the Milky Way},} \aap, 618, A93, \dodoi{10.1051/0004-6361/201833476}

\bibitem[{Z.~R. {Claytor} {et~al.}(2024){Claytor}, {van Saders}, {Cao}, {Pinsonneault}, {Teske}, \& {Beaton}}]{claytor2024}
{Claytor}, Z.~R., {van Saders}, J.~L., {Cao}, L., {et~al.} 2024, \bibinfo{title}{{TESS Stellar Rotation up to 80 Days in the Southern Continuous Viewing Zone},} \apj, 962, 47, \dodoi{10.3847/1538-4357/ad159a}

\bibitem[{G. {Clementini} {et~al.}(2023){Clementini}, {Ripepi}, {Garofalo}, {Molinaro}, {Muraveva}, {Leccia}, {Rimoldini}, {Holl}, {Jevardat de Fombelle}, {Sartoretti}, {Marchal}, {Audard}, {Nienartowicz}, {Andrae}, {Marconi}, {Szabados}, {Evans}, {Lecoeur-Taibi}, {Mowlavi}, {Musella}, \& {Eyer}}]{GaiaDR3_Variability_rrlyrae}
{Clementini}, G., {Ripepi}, V., {Garofalo}, A., {et~al.} 2023, \bibinfo{title}{{Gaia Data Release 3. Specific processing and validation of all-sky RR Lyrae and Cepheid stars: The RR Lyrae sample},} \aap, 674, A18, \dodoi{10.1051/0004-6361/202243964}

\bibitem[{I.~L. {Colman} {et~al.}(2024){Colman}, {Angus}, {David}, {Curtis}, {Hattori}, \& {Lu}}]{Colman2024}
{Colman}, I.~L., {Angus}, R., {David}, T., {et~al.} 2024, \bibinfo{title}{{Methods for the Detection of Stellar Rotation Periods in Individual TESS Sectors and Results from the Prime Mission},} \aj, 167, 189, \dodoi{10.3847/1538-3881/ad2c86}

\bibitem[{I.~L. {Colman} {et~al.}(2017){Colman}, {Huber}, {Bedding}, {Kuszlewicz}, {Yu}, {Beck}, {Elsworth}, {Garc{\'\i}a}, {Kawaler}, {Mathur}, {Stello}, \& {White}}]{Colman2017}
{Colman}, I.~L., {Huber}, D., {Bedding}, T.~R., {et~al.} 2017, \bibinfo{title}{{Evidence for compact binary systems around Kepler red giants},} \mnras, 469, 3802, \dodoi{10.1093/mnras/stx1056}

\bibitem[{K.~R. {Covey} {et~al.}(2016){Covey}, {Ag{\"u}eros}, {Law}, {Liu}, {Ahmadi}, {Laher}, {Levitan}, {Sesar}, \& {Surace}}]{covey2016}
{Covey}, K.~R., {Ag{\"u}eros}, M.~A., {Law}, N.~M., {et~al.} 2016, \bibinfo{title}{{Why Are Rapidly Rotating {{M}} Dwarfs in the {{Pleiades}} so (Infra)red? {{New}} Period Measurements Confirm Rotation-Dependent Color Offsets from the Cluster Sequence},} Astrophysical Journal, 822, 81, \dodoi{10.3847/0004-637X/822/2/81}

\bibitem[{J.~D. {Cummings} \& J.~S. {Kalirai}(2018){Cummings} \& {Kalirai}}]{cummings2018}
{Cummings}, J.~D., \& {Kalirai}, J.~S. 2018, \bibinfo{title}{{Improved Main-sequence Turnoff Ages of Young Open Clusters: Multicolor UBV Techniques and the Challenges of Rotation},} \aj, 156, 165, \dodoi{10.3847/1538-3881/aad5df}

\bibitem[{J.~L. {Curtis} {et~al.}(2019){Curtis}, {Ag{\"u}eros}, {Mamajek}, {Wright}, \& {Cummings}}]{curtis2019b}
{Curtis}, J.~L., {Ag{\"u}eros}, M.~A., {Mamajek}, E.~E., {Wright}, J.~T., \& {Cummings}, J.~D. 2019, \bibinfo{title}{{TESS Reveals that the Nearby Pisces-Eridanus Stellar Stream is only 120 Myr Old},} \aj, 158, 77, \dodoi{10.3847/1538-3881/ab2899}

\bibitem[{S.~T. {Douglas} {et~al.}(2019){Douglas}, {Curtis}, {Ag{\"u}eros}, {Cargile}, {Brewer}, {Meibom}, \& {Jansen}}]{douglas2019}
{Douglas}, S.~T., {Curtis}, J.~L., {Ag{\"u}eros}, M.~A., {et~al.} 2019, \bibinfo{title}{{K2 Rotation Periods for Low-mass Hyads and a Quantitative Comparison of the Distribution of Slow Rotators in the Hyades and Praesepe},} \apj, 879, 100, \dodoi{10.3847/1538-4357/ab2468}

\bibitem[{L. {Eyer} {et~al.}(2023){Eyer}, {Audard}, {Holl}, {Rimoldini}, {Carnerero}, {Clementini}, {De Ridder}, {Distefano}, {Evans}, {Gavras}, {Gomel}, {Lebzelter}, {Marton}, {Mowlavi}, {Panahi}, {Ripepi}, {Wyrzykowski}, {Nienartowicz}, {Jevardat de Fombelle}, {Lecoeur-Taibi}, {Rohrbasser}, {Riello}, {Garc{\'\i}a-Lario}, {Lanzafame}, {Mazeh}, {Raiteri}, {Zucker}, {{\'A}brah{\'a}m}, {Aerts}, {Aguado}, {Anderson}, {Bashi}, {Binnenfeld}, {Faigler}, {Garofalo}, {Karbevska}, {K{\'o}sp{\'a}l}, {Kruszy{\'n}ska}, {Kun}, {Lanza}, {Leccia}, {Marconi}, {Messina}, {Molinaro}, {Moln{\'a}r}, {Muraveva}, {Musella}, {Nagy}, {Pagano}, {Palaversa}, {Plachy}, {Pr{\v{s}}a}, {Rybicki}, {Shahaf}, {Szabados}, {Szegedi-Elek}, {Trabucchi}, {Barblan}, {Grenon}, {Roelens}, \& {S{\"u}veges}}]{GaiaDR3_Variability_method}
{Eyer}, L., {Audard}, M., {Holl}, B., {et~al.} 2023, \bibinfo{title}{{Gaia Data Release 3. Summary of the variability processing and analysis},} \aap, 674, A13, \dodoi{10.1051/0004-6361/202244242}

\bibitem[{A.~D. {Feinstein} {et~al.}(2019){Feinstein}, {Montet}, {Foreman-Mackey}, {Bedell}, {Saunders}, {Bean}, {Christiansen}, {Hedges}, {Luger}, {Scolnic}, \& {Cardoso}}]{Feinstein2019}
{Feinstein}, A.~D., {Montet}, B.~T., {Foreman-Mackey}, D., {et~al.} 2019, \bibinfo{title}{{eleanor: An Open-source Tool for Extracting Light Curves from the TESS Full-frame Images},} \pasp, 131, 094502, \dodoi{10.1088/1538-3873/ab291c}

\bibitem[{J.~A. {Fernandez} \& C.~W. {Salgado}(1980){Fernandez} \& {Salgado}}]{Fernandez1980}
{Fernandez}, J.~A., \& {Salgado}, C.~W. 1980, \bibinfo{title}{{Photometric study of the southern open cluster NGC 3532.},} \aaps, 39, 11

\bibitem[{A. {Frasca} {et~al.}(2023){Frasca}, {Alonso-Santiago}, {Catanzaro}, \& {Bragaglia}}]{frasca2023}
{Frasca}, A., {Alonso-Santiago}, J., {Catanzaro}, G., \& {Bragaglia}, A. 2023, \bibinfo{title}{{Rotation and activity in late-type members of the young cluster ASCC 123},} \mnras, 522, 4894, \dodoi{10.1093/mnras/stad1310}

\bibitem[{D.~J. {Fritzewski} {et~al.}(2021{\natexlab{a}}){Fritzewski}, {Barnes}, {James}, \& {\noopsort{a}{Strassmeier}, K.~G.}}]{fritz2021_periods}
{Fritzewski}, D.~J., {Barnes}, S.~A., {James}, D.~J., \& {\noopsort{a}{Strassmeier}, K.~G.} 2021{\natexlab{a}}, \bibinfo{title}{{Rotation periods for cool stars in the open cluster NGC 3532. The transition from fast to slow rotation},} \aap, 652, A60, \dodoi{10.1051/0004-6361/202140894}

\bibitem[{D.~J. {Fritzewski} {et~al.}(2021{\natexlab{b}}){Fritzewski}, {Barnes}, {James}, {\noopsort{b}{J{\"a}rvinen}, S.~P.}, \& {Strassmeier}}]{fritz2021_activity}
{Fritzewski}, D.~J., {Barnes}, S.~A., {James}, D.~J., {\noopsort{b}{J{\"a}rvinen}, S.~P.}, \& {Strassmeier}, K.~G. 2021{\natexlab{b}}, \bibinfo{title}{{A detailed understanding of the rotation-activity relationship using the 300 Myr old open cluster NGC 3532},} \aap, 656, A103, \dodoi{10.1051/0004-6361/202140896}

\bibitem[{D.~J. {Fritzewski} {et~al.}(2019){Fritzewski}, {Barnes}, {James}, {Geller}, {Meibom}, \& {Strassmeier}}]{fritz2019_rvs}
{Fritzewski}, D.~J., {Barnes}, S.~A., {James}, D.~J., {et~al.} 2019, \bibinfo{title}{{Spectroscopic membership for the populous 300 Myr-old open cluster NGC 3532},} \aap, 622, A110, \dodoi{10.1051/0004-6361/201833587}

\bibitem[{ {Gaia Collaboration} {et~al.}(2018){Gaia Collaboration}, {Babusiaux}, {van Leeuwen}, {Barstow}, {Jordi}, {Vallenari}, {Bossini}, {Bressan}, {Cantat-Gaudin}, {van Leeuwen}, {Brown}, {Prusti}, {de Bruijne}, {Bailer-Jones}, {Biermann}, {Evans}, {Eyer}, {Jansen}, {Klioner}, {Lammers}, {Lindegren}, {Luri}, {Mignard}, {Panem}, {Pourbaix}, {Randich}, {Sartoretti}, {Siddiqui}, {Soubiran}, {Walton}, {Arenou}, {Bastian}, {Cropper}, {Drimmel}, {Katz}, {Lattanzi}, {Bakker}, {Cacciari}, {Casta{\~n}eda}, {Chaoul}, {Cheek}, {De Angeli}, {Fabricius}, {Guerra}, {Holl}, {Masana}, {Messineo}, {Mowlavi}, {Nienartowicz}, {Panuzzo}, {Portell}, {Riello}, {Seabroke}, {Tanga}, {Th{\'e}venin}, {Gracia-Abril}, {Comoretto}, {Garcia-Reinaldos}, {Teyssier}, {Altmann}, {Andrae}, {Audard}, {Bellas-Velidis}, {Benson}, {Berthier}, {Blomme}, {Burgess}, {Busso}, {Carry}, {Cellino}, {Clementini}, {Clotet}, {Creevey}, {Davidson}, {De Ridder}, {Delchambre}, {Dell'Oro}, {Ducourant}, {Fern{\'a}ndez-Hern{\'a}ndez}, {Fouesneau},
  {Fr{\'e}mat}, {Galluccio}, {Garc{\'\i}a-Torres}, {Gonz{\'a}lez-N{\'u}{\~n}ez}, {Gonz{\'a}lez-Vidal}, {Gosset}, {Guy}, {Halbwachs}, {Hambly}, {Harrison}, {Hern{\'a}ndez}, {Hestroffer}, {Hodgkin}, {Hutton}, {Jasniewicz}, {Jean-Antoine-Piccolo}, {Jordan}, {Korn}, {Krone-Martins}, {Lanzafame}, {Lebzelter}, {L{\"o}ffler}, {Manteiga}, {Marrese}, {Mart{\'\i}n-Fleitas}, {Moitinho}, {Mora}, {Muinonen}, {Osinde}, {Pancino}, {Pauwels}, {Petit}, {Recio-Blanco}, {Richards}, {Rimoldini}, {Robin}, {Sarro}, {Siopis}, {Smith}, {Sozzetti}, {S{\"u}veges}, {Torra}, {van Reeven}, {Abbas}, {Abreu Aramburu}, {Accart}, {Aerts}, {Altavilla}, {{\'A}lvarez}, {Alvarez}, {Alves}, {Anderson}, {Andrei}, {Anglada Varela}, {Antiche}, {Antoja}, {Arcay}, {Astraatmadja}, {Bach}, {Baker}, {Balaguer-N{\'u}{\~n}ez}, {Balm}, {Barache}, {Barata}, {Barbato}, {Barblan}, {Barklem}, {Barrado}, {Barros}, {Bartholom{\'e} Mu{\~n}oz}, {Bassilana}, {Becciani}, {Bellazzini}, {Berihuete}, {Bertone}, {Bianchi}, {Bienaym{\'e}}, {Blanco-Cuaresma}, {Boch},
  {Boeche}, {Bombrun}, {Borrachero}, {Bouquillon}, {Bourda}, {Bragaglia}, {Bramante}, {Breddels}, {Brouillet}, {Br{\"u}semeister}, {Brugaletta}, {Bucciarelli}, {Burlacu}, {Busonero}, {Butkevich}, {Buzzi}, {Caffau}, {Cancelliere}, {Cannizzaro}, {Carballo}, {Carlucci}, {Carrasco}, {Casamiquela}, {Castellani}, {Castro-Ginard}, {Charlot}, {Chemin}, {Chiavassa}, {Cocozza}, {Costigan}, {Cowell}, {Crifo}, {Crosta}, {Crowley}, {Cuypers}, {Dafonte}, {Damerdji}, {Dapergolas}, {David}, {David}, {de Laverny}, {De Luise}, {De March}, {de Martino}, {de Souza}, {de Torres}, {Debosscher}, {del Pozo}, {Delbo}, {Delgado}, {Delgado}, {Diakite}, {Diener}, {Distefano}, {Dolding}, {Drazinos}, {Dur{\'a}n}, {Edvardsson}, {Enke}, {Eriksson}, {Esquej}, {Eynard Bontemps}, {Fabre}, {Fabrizio}, {Faigler}, {Falc{\~a}o}, {Farr{\`a}s Casas}, {Federici}, {Fedorets}, {Fernique}, {Figueras}, {Filippi}, {Findeisen}, {Fonti}, {Fraile}, {Fraser}, {Fr{\'e}zouls}, {Gai}, {Galleti}, {Garabato}, {Garc{\'\i}a-Sedano}, {Garofalo}, {Garralda}, {Gavel},
  {Gavras}, {Gerssen}, {Geyer}, {Giacobbe}, {Gilmore}, {Girona}, {Giuffrida}, {Glass}, {Gomes}, {Granvik}, {Gueguen}, {Guerrier}, {Guiraud}, {Guti{\'e}}, {Haigron}, {Hatzidimitriou}, {Hauser}, {Haywood}, {Heiter}, {Helmi}, {Heu}, {Hilger}, {Hobbs}, {Hofmann}, {Holland}, {Huckle}, {Hypki}, {Icardi}, {Jan{\ss}en}, {Jevardat de Fombelle}, {Jonker}, {Juh{\'a}sz}, {Julbe}, {Karampelas}, {Kewley}, {Klar}, {Kochoska}, {Kohley}, {Kolenberg}, {Kontizas}, {Kontizas}, {Koposov}, {Kordopatis}, {Kostrzewa-Rutkowska}, {Koubsky}, {Lambert}, {Lanza}, {Lasne}, {Lavigne}, {Le Fustec}, {Le Poncin-Lafitte}, {Lebreton}, {Leccia}, {Leclerc}, {Lecoeur-Taibi}, {Lenhardt}, {Leroux}, {Liao}, {Licata}, {Lindstr{\o}m}, {Lister}, {Livanou}, {Lobel}, {L{\'o}pez}, {Managau}, {Mann}, {Mantelet}, {Marchal}, {Marchant}, {Marconi}, {Marinoni}, {Marschalk{\'o}}, {Marshall}, {Martino}, {Marton}, {Mary}, {Massari}, {Matijevi{\v{c}}}, {Mazeh}, {McMillan}, {Messina}, {Michalik}, {Millar}, {Molina}, {Molinaro}, {Moln{\'a}r}, {Montegriffo}, {Mor},
  {Morbidelli}, {Morel}, {Morris}, {Mulone}, {Muraveva}, {Musella}, {Nelemans}, {Nicastro}, {Noval}, {O'Mullane}, {Ord{\'e}novic}, {Ord{\'o}{\~n}ez-Blanco}, {Osborne}, {Pagani}, {Pagano}, {Pailler}, {Palacin}, {Palaversa}, {Panahi}, {Pawlak}, {Piersimoni}, {Pineau}, {Plachy}, {Plum}, {Poggio}, {Poujoulet}, {Pr{\v{s}}a}, {Pulone}, {Racero}, {Ragaini}, {Rambaux}, {Ramos-Lerate}, {Regibo}, {Reyl{\'e}}, {Riclet}, {Ripepi}, {Riva}, {Rivard}, {Rixon}, {Roegiers}, {Roelens}, {Romero-G{\'o}mez}, {Rowell}, {Royer}, {Ruiz-Dern}, {Sadowski}, {Sagrist{\`a} Sell{\'e}s}, {Sahlmann}, {Salgado}, {Salguero}, {Sanna}, {Santana-Ros}, {Sarasso}, {Savietto}, {Schultheis}, {Sciacca}, {Segol}, {Segovia}, {S{\'e}gransan}, {Shih}, {Siltala}, {Silva}, {Smart}, {Smith}, {Solano}, {Solitro}, {Sordo}, {Soria Nieto}, {Souchay}, {Spagna}, {Spoto}, {Stampa}, {Steele}, {Steidelm{\"u}ller}, {Stephenson}, {Stoev}, {Suess}, {Surdej}, {Szabados}, {Szegedi-Elek}, {Tapiador}, {Taris}, {Tauran}, {Taylor}, {Teixeira}, {Terrett}, {Teyssandier},
  {Thuillot}, {Titarenko}, {Torra Clotet}, {Turon}, {Ulla}, {Utrilla}, {Uzzi}, {Vaillant}, {Valentini}, {Valette}, {van Elteren}, {Van Hemelryck}, {Vaschetto}, {Vecchiato}, {Veljanoski}, {Viala}, {Vicente}, {Vogt}, {von Essen}, {Voss}, {Votruba}, {Voutsinas}, {Walmsley}, {Weiler}, {Wertz}, {Wevers}, {Wyrzykowski}, {Yoldas}, {{\v{Z}}erjal}, {Ziaeepour}, {Zorec}, {Zschocke}, {Zucker}, {Zurbach}, \& {Zwitter}}]{gaia2018}
{Gaia Collaboration}, {Babusiaux}, C., {van Leeuwen}, F., {et~al.} 2018, \bibinfo{title}{{Gaia Data Release 2. Observational Hertzsprung-Russell diagrams},} \aap, 616, A10, \dodoi{10.1051/0004-6361/201832843}

\bibitem[{ {Gaia Collaboration} {et~al.}(2023){Gaia Collaboration}, {Vallenari}, {Brown}, {Prusti}, {de Bruijne}, {Arenou}, {Babusiaux}, {Biermann}, {Creevey}, {Ducourant}, {Evans}, {Eyer}, {Guerra}, {Hutton}, {Jordi}, {Klioner}, {Lammers}, {Lindegren}, {Luri}, {Mignard}, {Panem}, {Pourbaix}, {Randich}, {Sartoretti}, {Soubiran}, {Tanga}, {Walton}, {Bailer-Jones}, {Bastian}, {Drimmel}, {Jansen}, {Katz}, {Lattanzi}, {van Leeuwen}, {Bakker}, {Cacciari}, {Casta{\~n}eda}, {De Angeli}, {Fabricius}, {Fouesneau}, {Fr{\'e}mat}, {Galluccio}, {Guerrier}, {Heiter}, {Masana}, {Messineo}, {Mowlavi}, {Nicolas}, {Nienartowicz}, {Pailler}, {Panuzzo}, {Riclet}, {Roux}, {Seabroke}, {Sordo}, {Th{\'e}venin}, {Gracia-Abril}, {Portell}, {Teyssier}, {Altmann}, {Andrae}, {Audard}, {Bellas-Velidis}, {Benson}, {Berthier}, {Blomme}, {Burgess}, {Busonero}, {Busso}, {C{\'a}novas}, {Carry}, {Cellino}, {Cheek}, {Clementini}, {Damerdji}, {Davidson}, {de Teodoro}, {Nu{\~n}ez Campos}, {Delchambre}, {Dell'Oro}, {Esquej},
  {Fern{\'a}ndez-Hern{\'a}ndez}, {Fraile}, {Garabato}, {Garc{\'\i}a-Lario}, {Gosset}, {Haigron}, {Halbwachs}, {Hambly}, {Harrison}, {Hern{\'a}ndez}, {Hestroffer}, {Hodgkin}, {Holl}, {Jan{\ss}en}, {Jevardat de Fombelle}, {Jordan}, {Krone-Martins}, {Lanzafame}, {L{\"o}ffler}, {Marchal}, {Marrese}, {Moitinho}, {Muinonen}, {Osborne}, {Pancino}, {Pauwels}, {Recio-Blanco}, {Reyl{\'e}}, {Riello}, {Rimoldini}, {Roegiers}, {Rybizki}, {Sarro}, {Siopis}, {Smith}, {Sozzetti}, {Utrilla}, {van Leeuwen}, {Abbas}, {{\'A}brah{\'a}m}, {Abreu Aramburu}, {Aerts}, {Aguado}, {Ajaj}, {Aldea-Montero}, {Altavilla}, {{\'A}lvarez}, {Alves}, {Anders}, {Anderson}, {Anglada Varela}, {Antoja}, {Baines}, {Baker}, {Balaguer-N{\'u}{\~n}ez}, {Balbinot}, {Balog}, {Barache}, {Barbato}, {Barros}, {Barstow}, {Bartolom{\'e}}, {Bassilana}, {Bauchet}, {Becciani}, {Bellazzini}, {Berihuete}, {Bernet}, {Bertone}, {Bianchi}, {Binnenfeld}, {Blanco-Cuaresma}, {Blazere}, {Boch}, {Bombrun}, {Bossini}, {Bouquillon}, {Bragaglia}, {Bramante}, {Breedt},
  {Bressan}, {Brouillet}, {Brugaletta}, {Bucciarelli}, {Burlacu}, {Butkevich}, {Buzzi}, {Caffau}, {Cancelliere}, {Cantat-Gaudin}, {Carballo}, {Carlucci}, {Carnerero}, {Carrasco}, {Casamiquela}, {Castellani}, {Castro-Ginard}, {Chaoul}, {Charlot}, {Chemin}, {Chiaramida}, {Chiavassa}, {Chornay}, {Comoretto}, {Contursi}, {Cooper}, {Cornez}, {Cowell}, {Crifo}, {Cropper}, {Crosta}, {Crowley}, {Dafonte}, {Dapergolas}, {David}, {David}, {de Laverny}, {De Luise}, {De March}, {De Ridder}, {de Souza}, {de Torres}, {del Peloso}, {del Pozo}, {Delbo}, {Delgado}, {Delisle}, {Demouchy}, {Dharmawardena}, {Di Matteo}, {Diakite}, {Diener}, {Distefano}, {Dolding}, {Edvardsson}, {Enke}, {Fabre}, {Fabrizio}, {Faigler}, {Fedorets}, {Fernique}, {Fienga}, {Figueras}, {Fournier}, {Fouron}, {Fragkoudi}, {Gai}, {Garcia-Gutierrez}, {Garcia-Reinaldos}, {Garc{\'\i}a-Torres}, {Garofalo}, {Gavel}, {Gavras}, {Gerlach}, {Geyer}, {Giacobbe}, {Gilmore}, {Girona}, {Giuffrida}, {Gomel}, {Gomez}, {Gonz{\'a}lez-N{\'u}{\~n}ez},
  {Gonz{\'a}lez-Santamar{\'\i}a}, {Gonz{\'a}lez-Vidal}, {Granvik}, {Guillout}, {Guiraud}, {Guti{\'e}rrez-S{\'a}nchez}, {Guy}, {Hatzidimitriou}, {Hauser}, {Haywood}, {Helmer}, {Helmi}, {Sarmiento}, {Hidalgo}, {Hilger}, {H{\l}adczuk}, {Hobbs}, {Holland}, {Huckle}, {Jardine}, {Jasniewicz}, {Jean-Antoine Piccolo}, {Jim{\'e}nez-Arranz}, {Jorissen}, {Juaristi Campillo}, {Julbe}, {Karbevska}, {Kervella}, {Khanna}, {Kontizas}, {Kordopatis}, {Korn}, {K{\'o}sp{\'a}l}, {Kostrzewa-Rutkowska}, {Kruszy{\'n}ska}, {Kun}, {Laizeau}, {Lambert}, {Lanza}, {Lasne}, {Le Campion}, {Lebreton}, {Lebzelter}, {Leccia}, {Leclerc}, {Lecoeur-Taibi}, {Liao}, {Licata}, {Lindstr{\o}m}, {Lister}, {Livanou}, {Lobel}, {Lorca}, {Loup}, {Madrero Pardo}, {Magdaleno Romeo}, {Managau}, {Mann}, {Manteiga}, {Marchant}, {Marconi}, {Marcos}, {Marcos Santos}, {Mar{\'\i}n Pina}, {Marinoni}, {Marocco}, {Marshall}, {Martin Polo}, {Mart{\'\i}n-Fleitas}, {Marton}, {Mary}, {Masip}, {Massari}, {Mastrobuono-Battisti}, {Mazeh}, {McMillan}, {Messina}, {Michalik},
  {Millar}, {Mints}, {Molina}, {Molinaro}, {Moln{\'a}r}, {Monari}, {Mongui{\'o}}, {Montegriffo}, {Montero}, {Mor}, {Mora}, {Morbidelli}, {Morel}, {Morris}, {Muraveva}, {Murphy}, {Musella}, {Nagy}, {Noval}, {Oca{\~n}a}, {Ogden}, {Ordenovic}, {Osinde}, {Pagani}, {Pagano}, {Palaversa}, {Palicio}, {Pallas-Quintela}, {Panahi}, {Payne-Wardenaar}, {Pe{\~n}alosa Esteller}, {Penttil{\"a}}, {Pichon}, {Piersimoni}, {Pineau}, {Plachy}, {Plum}, {Poggio}, {Pr{\v{s}}a}, {Pulone}, {Racero}, {Ragaini}, {Rainer}, {Raiteri}, {Rambaux}, {Ramos}, {Ramos-Lerate}, {Re Fiorentin}, {Regibo}, {Richards}, {Rios Diaz}, {Ripepi}, {Riva}, {Rix}, {Rixon}, {Robichon}, {Robin}, {Robin}, {Roelens}, {Rogues}, {Rohrbasser}, {Romero-G{\'o}mez}, {Rowell}, {Royer}, {Ruz Mieres}, {Rybicki}, {Sadowski}, {S{\'a}ez N{\'u}{\~n}ez}, {Sagrist{\`a} Sell{\'e}s}, {Sahlmann}, {Salguero}, {Samaras}, {Sanchez Gimenez}, {Sanna}, {Santove{\~n}a}, {Sarasso}, {Schultheis}, {Sciacca}, {Segol}, {Segovia}, {S{\'e}gransan}, {Semeux}, {Shahaf}, {Siddiqui}, {Siebert},
  {Siltala}, {Silvelo}, {Slezak}, {Slezak}, {Smart}, {Snaith}, {Solano}, {Solitro}, {Souami}, {Souchay}, {Spagna}, {Spina}, {Spoto}, {Steele}, {Steidelm{\"u}ller}, {Stephenson}, {S{\"u}veges}, {Surdej}, {Szabados}, {Szegedi-Elek}, {Taris}, {Taylor}, {Teixeira}, {Tolomei}, {Tonello}, {Torra}, {Torra}, {Torralba Elipe}, {Trabucchi}, {Tsounis}, {Turon}, {Ulla}, {Unger}, {Vaillant}, {van Dillen}, {van Reeven}, {Vanel}, {Vecchiato}, {Viala}, {Vicente}, {Voutsinas}, {Weiler}, {Wevers}, {Wyrzykowski}, {Yoldas}, {Yvard}, {Zhao}, {Zorec}, {Zucker}, \& {Zwitter}}]{gaiadr3}
{Gaia Collaboration}, {Vallenari}, A., {Brown}, A.~G.~A., {et~al.} 2023, \bibinfo{title}{{Gaia Data Release 3. Summary of the content and survey properties},} \aap, 674, A1, \dodoi{10.1051/0004-6361/202243940}

\bibitem[{A. {Ginsburg} {et~al.}(2019){Ginsburg}, {Sip{\H{o}}cz}, {Brasseur}, {Cowperthwaite}, {Craig}, {Deil}, {Guillochon}, {Guzman}, {Liedtke}, {Lian Lim}, {Lockhart}, {Mommert}, {Morris}, {Norman}, {Parikh}, {Persson}, {Robitaille}, {Segovia}, {Singer}, {Tollerud}, {de Val-Borro}, {Valtchanov}, {Woillez}, {Astroquery Collaboration}, \& {a subset of astropy Collaboration}}]{astroquery}
{Ginsburg}, A., {Sip{\H{o}}cz}, B.~M., {Brasseur}, C.~E., {et~al.} 2019, \bibinfo{title}{{astroquery: An Astronomical Web-querying Package in Python},} \aj, 157, 98, \dodoi{10.3847/1538-3881/aafc33}

\bibitem[{T. {Han} \& T.~D. {Brandt}(2023){Han} \& {Brandt}}]{TGLC}
{Han}, T., \& {Brandt}, T.~D. 2023, \bibinfo{title}{{TESS-Gaia Light Curve: A PSF-based TESS FFI Light-curve Product},} \aj, 165, 71, \dodoi{10.3847/1538-3881/acaaa7}

\bibitem[{C.~R. Harris {et~al.}(2020)Harris, Millman, van~der Walt, Gommers, Virtanen, Cournapeau, Wieser, Taylor, Berg, Smith, Kern, Picus, Hoyer, van Kerkwijk, Brett, Haldane, Fernández~del Río, Wiebe, Peterson, Gérard-Marchant, Sheppard, Reddy, Weckesser, Abbasi, Gohlke, \& Oliphant}]{Numpy2020}
Harris, C.~R., Millman, K.~J., van~der Walt, S.~J., {et~al.} 2020, \bibinfo{title}{Array programming with {NumPy},} Nature, 585, 357–362, \dodoi{10.1038/s41586-020-2649-2}

\bibitem[{S. {Hattori} {et~al.}(2025){Hattori}, {Angus}, {Foreman-Mackey}, {Lu}, \& {Colman}}]{hattori2025}
{Hattori}, S., {Angus}, R., {Foreman-Mackey}, D., {Lu}, Y.~L., \& {Colman}, I. 2025, \bibinfo{title}{{Measuring Long Stellar Rotation Periods (>10 days) from TESS FFI Light Curves is Possible: An Investigation Using TESS and ZTF},} \aj, 170, 15, \dodoi{10.3847/1538-3881/add0ab}

\bibitem[{S. {Hattori} {et~al.}(2022){Hattori}, {Foreman-Mackey}, {Hogg}, {Montet}, {Angus}, {Pritchard}, {Curtis}, \& {Sch{\"o}lkopf}}]{Hattori2022}
{Hattori}, S., {Foreman-Mackey}, D., {Hogg}, D.~W., {et~al.} 2022, \bibinfo{title}{{The unpopular Package: A Data-driven Approach to Detrending TESS Full-frame Image Light Curves},} \aj, 163, 284, \dodoi{10.3847/1538-3881/ac625a}

\bibitem[{S. {Herschel}(1847){Herschel}}]{herschel1847}
{Herschel}, John Frederick~William, S. 1847, {Results of astronomical observations made during the years 1834, 5, 6, 7, 8, at the Cape of Good Hope; being the completion of a telescopic survey of the whole surface of the visible heavens, commenced in 1825}

\bibitem[{M.~E. {Higgins} \& K.~J. {Bell}(2023){Higgins} \& {Bell}}]{Higgins2023}
{Higgins}, M.~E., \& {Bell}, K.~J. 2023, \bibinfo{title}{{Localizing Sources of Variability in Crowded TESS Photometry},} \aj, 165, 141, \dodoi{10.3847/1538-3881/acb20c}

\bibitem[{S.~B. Howell {et~al.}(2014)Howell, Sobeck, Haas, Still, Barclay, Mullally, Troeltzsch, Aigrain, Bryson, Caldwell, Chaplin, Cochran, Huber, Marcy, Miglio, Najita, Smith, Twicken, \& Fortney}]{howell2014}
Howell, S.~B., Sobeck, C., Haas, M., {et~al.} 2014, \bibinfo{title}{The {{K2 Mission}}: {{Characterization}} and {{Early Results}},} Publications of the Astronomical Society of the Pacific, 126, 398, \dodoi{10.1086/676406}

\bibitem[{E.~L. {Hunt} \& S. {Reffert}(2021){Hunt} \& {Reffert}}]{Hunt2021}
{Hunt}, E.~L., \& {Reffert}, S. 2021, \bibinfo{title}{{Improving the open cluster census. I. Comparison of clustering algorithms applied to Gaia DR2 data},} \aap, 646, A104, \dodoi{10.1051/0004-6361/202039341}

\bibitem[{E.~L. {Hunt} \& S. {Reffert}(2023){Hunt} \& {Reffert}}]{Hunt2023}
{Hunt}, E.~L., \& {Reffert}, S. 2023, \bibinfo{title}{{Improving the open cluster census. II. An all-sky cluster catalogue with Gaia DR3},} \aap, 673, A114, \dodoi{10.1051/0004-6361/202346285}

\bibitem[{J.~D. {Hunter}(2007){Hunter}}]{matplotlib}
{Hunter}, J.~D. 2007, \bibinfo{title}{{Matplotlib: A 2D Graphics Environment},} Computing in Science and Engineering, 9, 90, \dodoi{10.1109/MCSE.2007.55}

\bibitem[{E. Jones {et~al.}(2001)Jones, Oliphant, Peterson, {et~al.}}]{scipy}
Jones, E., Oliphant, T., Peterson, P., {et~al.} 2001, \bibinfo{title}{{SciPy}: Open source scientific tools for {Python},} \url{http://www.scipy.org/}

\bibitem[{D. {Koelbloed}(1959){Koelbloed}}]{Koelbloed1959}
{Koelbloed}, D. 1959, \bibinfo{title}{{Three-colour photometry of the three southern open clusters NGC 3532, 6475 (M7) and 6124 .},} \bain, 14, 265

\bibitem[{M. {Kounkel} \& K. {Covey}(2019){Kounkel} \& {Covey}}]{kounkel2019}
{Kounkel}, M., \& {Covey}, K. 2019, \bibinfo{title}{{Untangling the Galaxy. I. Local Structure and Star Formation History of the Milky Way},} \aj, 158, 122, \dodoi{10.3847/1538-3881/ab339a}

\bibitem[{M. {Kounkel} {et~al.}(2020){Kounkel}, {Covey}, \& {Stassun}}]{kounkel2020}
{Kounkel}, M., {Covey}, K., \& {Stassun}, K.~G. 2020, \bibinfo{title}{{Untangling the Galaxy. II. Structure within 3 kpc},} \aj, 160, 279, \dodoi{10.3847/1538-3881/abc0e6}

\bibitem[{ {Lightkurve Collaboration} {et~al.}(2018){Lightkurve Collaboration}, {Cardoso}, {Hedges}, {Gully-Santiago}, {Saunders}, {Cody}, {Barclay}, {Hall}, {Sagear}, {Turtelboom}, {Zhang}, {Tzanidakis}, {Mighell}, {Coughlin}, {Bell}, {Berta-Thompson}, {Williams}, {Dotson}, \& {Barentsen}}]{lightkurve}
{Lightkurve Collaboration}, {Cardoso}, J.~V.~d.~M., {Hedges}, C., {et~al.} 2018, \bibinfo{title}{{Lightkurve: Kepler and TESS time series analysis in Python},}, Astrophysics Source Code Library \doeprint{1812.013}

\bibitem[{N.~R. {Lomb}(1976){Lomb}}]{Lomb1976}
{Lomb}, N.~R. 1976, \bibinfo{title}{{Least-Squares Frequency Analysis of Unequally Spaced Data},} \apss, 39, 447, \dodoi{10.1007/BF00648343}

\bibitem[{L. {McInnes} {et~al.}(2017){McInnes}, {Healy}, \& {Astels}}]{hdbscan2017}
{McInnes}, L., {Healy}, J., \& {Astels}, S. 2017, \bibinfo{title}{{hdbscan: Hierarchical density based clustering},} The Journal of Open Source Software, 2, 205, \dodoi{10.21105/joss.00205}

\bibitem[{N. {Mowlavi} {et~al.}(2023){Mowlavi}, {Holl}, {Lecoeur-Ta{\"\i}bi}, {Barblan}, {Kochoska}, {Pr{\v{s}}a}, {Mazeh}, {Rimoldini}, {Gavras}, {Audard}, {Jevardat de Fombelle}, {Nienartowicz}, {Garc{\'\i}a-Lario}, \& {Eyer}}]{GaiaDR3_Variability_ebs}
{Mowlavi}, N., {Holl}, B., {Lecoeur-Ta{\"\i}bi}, I., {et~al.} 2023, \bibinfo{title}{{Gaia Data Release 3. The first Gaia catalogue of eclipsing-binary candidates},} \aap, 674, A16, \dodoi{10.1051/0004-6361/202245330}

\bibitem[{D. {Nardiello} {et~al.}(2019){Nardiello}, {Borsato}, {Piotto}, {Colombo}, {Manthopoulou}, {Bedin}, {Granata}, {Lacedelli}, {Libralato}, {Malavolta}, {Montalto}, \& {Nascimbeni}}]{PATHOS}
{Nardiello}, D., {Borsato}, L., {Piotto}, G., {et~al.} 2019, \bibinfo{title}{{A PSF-based Approach to TESS High quality data Of Stellar clusters (PATHOS) - I. Search for exoplanets and variable stars in the field of 47 Tuc},} \mnras, 490, 3806, \dodoi{10.1093/mnras/stz2878}

\bibitem[{E.~R. {Newton} {et~al.}(2022){Newton}, {Rampalli}, {Kraus}, {Mann}, {Curtis}, {Vanderburg}, {Krolikowski}, {Huber}, {Petter}, {Bieryla}, {Tofflemire}, {Thao}, {Wood}, {Kerr}, {Safanov}, {Strakhov}, {Ciardi}, {Giacalone}, {Dressing}, {Gill}, {Savel}, {Collins}, {Brown}, {Murgas}, {Isogai}, {Narita}, {Palle}, {Quinn}, {Eastman}, {F{\H{u}}r{\'e}sz}, {Shiao}, {Daylan}, {Caldwell}, {Ricker}, {Vanderspek}, {Seager}, {Winn}, {Jenkins}, \& {Latham}}]{newton2022}
{Newton}, E.~R., {Rampalli}, R., {Kraus}, A.~L., {et~al.} 2022, \bibinfo{title}{{TESS Hunt for Young and Maturing Exoplanets (THYME). VII. Membership, Rotation, and Lithium in the Young Cluster Group-X and a New Young Exoplanet},} \aj, 164, 115, \dodoi{10.3847/1538-3881/ac8154}

\bibitem[{F. {Ochsenbein} {et~al.}(2000){Ochsenbein}, {Bauer}, \& {Marcout}}]{vizier}
{Ochsenbein}, F., {Bauer}, P., \& {Marcout}, J. 2000, \bibinfo{title}{{The VizieR database of astronomical catalogues},} \aaps, 143, 23, \dodoi{10.1051/aas:2000169}

\bibitem[{R.~J. {Oelkers} \& K.~G. {Stassun}(2018){Oelkers} \& {Stassun}}]{Oelkers2018}
{Oelkers}, R.~J., \& {Stassun}, K.~G. 2018, \bibinfo{title}{{Precision Light Curves from TESS Full-frame Images: A Different Imaging Approach},} \aj, 156, 132, \dodoi{10.3847/1538-3881/aad68e}

\bibitem[{E.~K. {Pass} {et~al.}(2022){Pass}, {Charbonneau}, {Irwin}, \& {Winters}}]{pass2022}
{Pass}, E.~K., {Charbonneau}, D., {Irwin}, J.~M., \& {Winters}, J.~G. 2022, \bibinfo{title}{{Constraints on the Spindown of Fully Convective M Dwarfs Using Wide Field Binaries},} \apj, 936, 109, \dodoi{10.3847/1538-4357/ac7da8}

\bibitem[{R.~P. {Petrucci} {et~al.}(2024){Petrucci}, {G{\'o}mez Maqueo Chew}, {Jofr{\'e}}, {Segura}, \& {Ferrero}}]{petrucci2024}
{Petrucci}, R.~P., {G{\'o}mez Maqueo Chew}, Y., {Jofr{\'e}}, E., {Segura}, A., \& {Ferrero}, L.~V. 2024, \bibinfo{title}{{Exploring the photometric variability of ultra-cool dwarfs with TESS},} \mnras, 527, 8290, \dodoi{10.1093/mnras/stad3720}

\bibitem[{M. {Popinchalk} {et~al.}(2023){Popinchalk}, {Faherty}, {Curtis}, {Gagn{\'e}}, {Bardalez Gagliuffi}, {Vos}, {Ayala}, {Gonzales}, \& {Kiman}}]{popinchalk2023}
{Popinchalk}, M., {Faherty}, J.~K., {Curtis}, J.~L., {et~al.} 2023, \bibinfo{title}{{Examining the Rotation Period Distribution of the 40 Myr Tucana-Horologium Association with TESS},} \apj, 945, 114, \dodoi{10.3847/1538-4357/acb055}

\bibitem[{W.~H. Press \& G.~B. Rybicki(1989)Press \& Rybicki}]{press1989}
Press, W.~H., \& Rybicki, G.~B. 1989, \bibinfo{title}{Fast Algorithm for Spectral Analysis of Unevenly Sampled Data,} Astrophysical Journal, 338, 277, \dodoi{10.1086/167197}

\bibitem[{R. {Rampalli} {et~al.}(2021){Rampalli}, {Ag{\"u}eros}, {Curtis}, {Douglas}, {N{\'u}{\~n}ez}, {Cargile}, {Covey}, {Gosnell}, {Kraus}, {Law}, \& {Mann}}]{Rampalli2021}
{Rampalli}, R., {Ag{\"u}eros}, M.~A., {Curtis}, J.~L., {et~al.} 2021, \bibinfo{title}{{Three K2 Campaigns Yield Rotation Periods for 1013 Stars in Praesepe},} \apj, 921, 167, \dodoi{10.3847/1538-4357/ac0c1e}

\bibitem[{L.~M. {Rebull} {et~al.}(2022){Rebull}, {Stauffer}, {Hillenbrand}, {Cody}, {Kruse}, \& {Powell}}]{rebull2022}
{Rebull}, L.~M., {Stauffer}, J.~R., {Hillenbrand}, L.~A., {et~al.} 2022, \bibinfo{title}{{Rotation of Low-mass Stars in Upper Centaurus-Lupus and Lower Centaurus-Crux with TESS},} \aj, 164, 80, \dodoi{10.3847/1538-3881/ac75f1}

\bibitem[{L.~M. Rebull {et~al.}(2016)Rebull, Stauffer, Bouvier, Cody, Hillenbrand, Soderblom, Valenti, Barrado, Bouy, Ciardi, Pinsonneault, Stassun, Micela, Aigrain, Vrba, Somers, Christiansen, Gillen, \& Collier~Cameron}]{rebull2016}
Rebull, L.~M., Stauffer, J.~R., Bouvier, J., {et~al.} 2016, \bibinfo{title}{Rotation in the {{Pleiades}} with {{K2}}. {{I}}. {{Data}} and {{First Results}},} AJ, 152, 113, \dodoi{10.3847/0004-6256/152/5/113}

\bibitem[{G.~R. {Ricker} {et~al.}(2015){Ricker}, {Winn}, {Vanderspek}, {Latham}, {Bakos}, {Bean}, {Berta-Thompson}, {Brown}, {Buchhave}, {Butler}, {Butler}, {Chaplin}, {Charbonneau}, {Christensen-Dalsgaard}, {Clampin}, {Deming}, {Doty}, {De Lee}, {Dressing}, {Dunham}, {Endl}, {Fressin}, {Ge}, {Henning}, {Holman}, {Howard}, {Ida}, {Jenkins}, {Jernigan}, {Johnson}, {Kaltenegger}, {Kawai}, {Kjeldsen}, {Laughlin}, {Levine}, {Lin}, {Lissauer}, {MacQueen}, {Marcy}, {McCullough}, {Morton}, {Narita}, {Paegert}, {Palle}, {Pepe}, {Pepper}, {Quirrenbach}, {Rinehart}, {Sasselov}, {Sato}, {Seager}, {Sozzetti}, {Stassun}, {Sullivan}, {Szentgyorgyi}, {Torres}, {Udry}, \& {Villasenor}}]{Ricker2015}
{Ricker}, G.~R., {Winn}, J.~N., {Vanderspek}, R., {et~al.} 2015, \bibinfo{title}{{Transiting Exoplanet Survey Satellite (TESS)},} J.~Astron.~Telesc.~Instrum.~Syst., 1, 014003, \dodoi{10.1117/1.JATIS.1.1.014003}

\bibitem[{V. {Ripepi} {et~al.}(2023){Ripepi}, {Clementini}, {Molinaro}, {Leccia}, {Plachy}, {Moln{\'a}r}, {Rimoldini}, {Musella}, {Marconi}, {Garofalo}, {Audard}, {Holl}, {Evans}, {Jevardat de Fombelle}, {Lecoeur-Taibi}, {Marchal}, {Mowlavi}, {Muraveva}, {Nienartowicz}, {Sartoretti}, {Szabados}, \& {Eyer}}]{GaiaDR3_Variability_cepheids}
{Ripepi}, V., {Clementini}, G., {Molinaro}, R., {et~al.} 2023, \bibinfo{title}{{Gaia Data Release 3. Specific processing and validation of all sky RR Lyrae and Cepheid stars: The Cepheid sample},} \aap, 674, A17, \dodoi{10.1051/0004-6361/202243990}

\bibitem[{M. {Roelens} {et~al.}(2018){Roelens}, {Eyer}, {Mowlavi}, {Rimoldini}, {Lecoeur-Ta{\"\i}bi}, {Nienartowicz}, {Jevardat de Fombelle}, {Marchal}, {Audard}, {Guy}, {Holl}, {Evans}, {Riello}, {De Angeli}, {Blanco-Cuaresma}, \& {Wevers}}]{GaiaDR2_Variability_short}
{Roelens}, M., {Eyer}, L., {Mowlavi}, N., {et~al.} 2018, \bibinfo{title}{{Gaia Data Release 2. Short-timescale variability processing and analysis},} \aap, 620, A197, \dodoi{10.1051/0004-6361/201833357}

\bibitem[{J.~D. {Scargle}(1982){Scargle}}]{Scargle1982}
{Scargle}, J.~D. 1982, \bibinfo{title}{{Studies in astronomical time series analysis. II - Statistical aspects of spectral analysis of unevenly spaced data},} \apj, 263, 835, \dodoi{10.1086/160554}

\bibitem[{ STScI(2022)STScI}]{TESSdata}
STScI. 2022, \bibinfo{title}{TESS Calibrated Full Frame Images: All Sectors,} STScI/MAST, \dodoi{10.17909/0CP4-2J79}

\bibitem[{ {The Astropy Collaboration} {et~al.}(2018){The Astropy Collaboration}, {Price-Whelan}, {Sip{\H o}cz}, {G{\"u}nther}, {Lim}, {Crawford}, {Conseil}, {Shupe}, {Craig}, {Dencheva}, {Ginsburg}, {VanderPlas}, {Bradley}, {P{\'e}rez-Su{\'a}rez}, {de Val-Borro}, {Aldcroft}, {Cruz}, {Robitaille}, {Tollerud}, {Ardelean}, {Babej}, {Bachetti}, {Bakanov}, {Bamford}, {Barentsen}, {Barmby}, {Baumbach}, {Berry}, {Biscani}, {Boquien}, {Bostroem}, {Bouma}, {Brammer}, {Bray}, {Breytenbach}, {Buddelmeijer}, {Burke}, {Calderone}, {Cano Rodr{\'{\i}}guez}, {Cara}, {Cardoso}, {Cheedella}, {Copin}, {Crichton}, {D{\'A}vella}, {Deil}, {Depagne}, {Dietrich}, {Donath}, {Droettboom}, {Earl}, {Erben}, {Fabbro}, {Ferreira}, {Finethy}, {Fox}, {Garrison}, {Gibbons}, {Goldstein}, {Gommers}, {Greco}, {Greenfield}, {Groener}, {Grollier}, {Hagen}, {Hirst}, {Homeier}, {Horton}, {Hosseinzadeh}, {Hu}, {Hunkeler}, {Ivezi{\'c}}, {Jain}, {Jenness}, {Kanarek}, {Kendrew}, {Kern}, {Kerzendorf}, {Khvalko}, {King}, {Kirkby}, {Kulkarni}, {Kumar},
  {Lee}, {Lenz}, {Littlefair}, {Ma}, {Macleod}, {Mastropietro}, {McCully}, {Montagnac}, {Morris}, {Mueller}, {Mumford}, {Muna}, {Murphy}, {Nelson}, {Nguyen}, {Ninan}, {N{\"o}the}, {Ogaz}, {Oh}, {Parejko}, {Parley}, {Pascual}, {Patil}, {Patil}, {Plunkett}, {Prochaska}, {Rastogi}, {Reddy Janga}, {Sabater}, {Sakurikar}, {Seifert}, {Sherbert}, {Sherwood-Taylor}, {Shih}, {Sick}, {Silbiger}, {Singanamalla}, {Singer}, {Sladen}, {Sooley}, {Sornarajah}, {Streicher}, {Teuben}, {Thomas}, {Tremblay}, {Turner}, {Terr{\'o}n}, {van Kerkwijk}, {de la Vega}, {Watkins}, {Weaver}, {Whitmore}, {Woillez}, \& {Zabalza}}]{Astropy2018}
{The Astropy Collaboration}, {Price-Whelan}, A.~M., {Sip{\H o}cz}, B.~M., {et~al.} 2018, \bibinfo{title}{{The Astropy Project: Building an inclusive, open-science project and status of the v2.0 core package},} ArXiv e-prints.
\newblock \doarXiv{1801.02634}

\bibitem[{P.~R. {Van-Lane} {et~al.}(2025){Van-Lane}, {Speagle}, {Eadie}, {Douglas}, {Cargile}, {Zucker}, {Lu}, \& {Angus}}]{ChronoFlow2025}
{Van-Lane}, P.~R., {Speagle}, J.~S., {Eadie}, G.~M., {et~al.} 2025, \bibinfo{title}{{ChronoFlow: A Data-driven Model for Gyrochronology},} \apj, 986, 59, \dodoi{10.3847/1538-4357/adcd73}

\bibitem[{D. {Wang} {et~al.}(2016){Wang}, {Hogg}, {Foreman-Mackey}, \& {Sch{\"o}lkopf}}]{wang2016}
{Wang}, D., {Hogg}, D.~W., {Foreman-Mackey}, D., \& {Sch{\"o}lkopf}, B. 2016, \bibinfo{title}{{A Causal, Data-driven Approach to Modeling the Kepler Data},} \pasp, 128, 094503, \dodoi{10.1088/1538-3873/128/967/094503}

\bibitem[{M. {Wenger} {et~al.}(2000){Wenger}, {Ochsenbein}, {Egret}, {Dubois}, {Bonnarel}, {Borde}, {Genova}, {Jasniewicz}, {Lalo{\"e}}, {Lesteven}, \& {Monier}}]{simbad}
{Wenger}, M., {Ochsenbein}, F., {Egret}, D., {et~al.} 2000, \bibinfo{title}{{The SIMBAD astronomical database. The CDS reference database for astronomical objects},} \aaps, 143, 9, \dodoi{10.1051/aas:2000332}

\end{thebibliography}
\bibliographystyle{aasjournalv7}

{\providecommand{\noopsort}[1]{}}\def\noopsort#1{}

%% This command is needed to show the entire author+affiliation list when
%% the collaboration and author truncation commands are used.  It has to
%% go at the end of the manuscript.
%\allauthors

%% Include this line if you are using the \added, \replaced, \deleted
%% commands to see a summary list of all changes at the end of the article.
%\listofchanges

\end{document}